\documentclass[11pt]{article}
\pdfoutput=1
\usepackage{graphicx}
\usepackage{amsmath,amssymb,amsthm,amsxtra,overpic,bbm,bm,epsfig,ulem}
\usepackage{color}
\usepackage{multirow}
\usepackage{rotating}
\usepackage{slashbox}

\usepackage{array}
\newcommand{\PreserveBackslash}[1]{\let\temp=\\#1\let\\=\temp}
\newcolumntype{C}[1]{>{\PreserveBackslash\centering}p{#1}}
\newcolumntype{R}[1]{>{\PreserveBackslash\raggedleft}p{#1}}
\newcolumntype{L}[1]{>{\PreserveBackslash\raggedright}p{#1}}

\makeatletter

\newcommand{\Rmnum}[1]{\expandafter\@slowromancap\romannumeral #1@}
\makeatother

\begin{document}

\begin{center}
{\Large \bf Breakings of the neutrino $\mu$-$\tau$ reflection symmetry}
\end{center}

\vspace{0.1cm}

\begin{center}
{\bf Zhen-hua Zhao} \footnote{E-mail: zhzhao@itp.ac.cn} \\
{ Department of Physics, Liaoning Normal University, Dalian 116029, China}
\end{center}

\vspace{0.5cm}

\begin{abstract}
The neutrino $\mu$-$\tau$ reflection symmetry has been attracting a lot of attention as
it predicts the interesting results $\theta^{}_{23} = \pi/4$ and $\delta = \pm \pi/2$.
But it is reasonable to consider breakings of such a symmetry either from the theoretical
considerations or on the basis of experimental results. We thus perform a systematic
study for the possible symmetry-breaking patterns and their implications for the
mixing parameters. The general treatment is applied to some specific symmetry
breaking arising from the renormalization group effects for illustration.
\end{abstract}

\newpage

\section{Introduction}

The discovery of neutrino oscillations indicates that neutrinos are massive and bear
flavor mixing. The neutrino mixing arises from the mismatch between their mass and flavor
eigenstates, and is described by a $3 \times 3$ unitary matrix $U= U^\dagger_l U^{}_\nu$
(with $U^{}_l$ and $U^{}_\nu$ being respectively the unitary matrix for diagonalizing the
charged-lepton mass matrix $M^{}_l M^\dagger_l$ and neutrino mass matrix $M^{}_\nu$).
In the standard parametrization, $U$ reads
\begin{equation}
U = P^{}_\phi \left( \begin{matrix}
c_{12} c_{13} & s_{12} c_{13} & s_{13} e^{-{\rm i} \delta} \cr
-s_{12} c_{23} - c_{12} s_{23} s_{13} e^{{\rm i} \delta}
& c_{12} c_{23} - s_{12} s_{23} s_{13} e^{{\rm i} \delta}  & s_{23} c_{13} \cr
s_{12} s_{23} - c_{12} c_{23} s_{13} e^{{\rm i} \delta}
& -c_{12} s_{23} - s_{12} c_{23} s_{13} e^{{\rm i} \delta} & c_{23}c_{13}
\end{matrix} \right) P^{}_\nu \;,
\label{1.1}
\end{equation}
where $\theta^{}_{ij}$ (for $ij=12, 13, 23$) are the mixing angles
(with $c^{}_{ij} = \cos{\theta^{}_{ij} }$ and $s^{}_{ij} = \sin {\theta^{}_{ij}}$) and
$\delta$ is the Dirac CP phase.
$P^{}_\nu = {\rm Diag}(e^{{\rm i} \rho}, e^{{\rm i} \sigma}, 1)$ contains two Majorana CP
phases $\rho$ and $\sigma$, while $P^{}_{\phi} = {\rm Diag}(e^{{\rm i} \phi^{}_1}, e^{{\rm i} \phi^{}_2},
e^{{\rm i} \phi^{}_3})$ consists of three unphysical phases $\phi^{}_{1, 2, 3}$ that can be removed via the
charged-lepton field rephasing. In addition, neutrino oscillations are also controlled by
two mass-squared differences $\Delta m^2_{ij} = m^2_i - m^2_j$ (for $ij = 21, 31$). Thanks
to various neutrino-oscillation experiments \cite{pdg}, the neutrino mixing parameters have
been measured to a good accuracy. A global-fit result \cite{global} for them is given by
\begin{eqnarray}
&& \sin^2{\theta^{}_{12}} = 0.308 \pm 0.017 \;, \hspace{1.6cm}
\Delta m^2_{21}= (7.54 \pm 0.24) \times 10^{-5} \ {\rm eV}^2 \;, \nonumber \\
&& \sin^2{\theta^{}_{13}} = 0.0234 \pm 0.0020 \;,\hspace{1cm}
|\Delta m^2_{31}| = (2.47 \pm 0.06) \times 10^{-3} \ {\rm eV}^2¡¡\;.
\label{1.2}
\end{eqnarray}
Note that the sign of $\Delta m^2_{31}$ remains undetermined, allowing for two possible
neutrino mass orderings $m^{}_1 < m^{}_2 < m^{}_3$ (referred to as the normal hierarchy and NH
for short) or $m^{}_3 < m^{}_1 < m^{}_2$ (the inverted hierarchy and IH for short).
The absolute neutrino mass scale is not known either, but subject to the constraint $m^{}_1 +
m^{}_2 + m^{}_3 < 0.23$ eV from cosmological observations \cite{planck}. Particularly
noteworthy, a recent result from the NOvA experiment ($\theta^{}_{23} = 39.5^\circ \pm 1.7^\circ$
or $52.1^\circ \pm 1.7^\circ$ in the NH case) disfavors the popular maximal mixing scenario
$\theta^{}_{23}= 45^\circ$ with 2.6$\sigma$ significance \cite{NOvA}. On the other hand, it is
interesting to find that the best-fit result for $\delta$ is around $270^\circ$ ($261^\circ \pm
55^\circ$ for NH and $277^\circ \pm 43^\circ$ for IH) \cite{global2}.

How to understand the neutrino mixing pattern poses an interesting question. As symmetries
(e.g., the $\rm SU(3)^{}_q$ quark flavor symmetry) have been serving as a guideline for understanding
the particle physics, they may play a similar role in addressing the flavor issues. Along this line,
many discrete groups have been proposed as the lepton flavor
symmetry \cite{review}. A simplest example is the $\mu$-$\tau$ permutation symmetry
\cite{MT,review2}: In the basis of $M^{}_l$ being diagonal,
$M^{}_\nu$ should keep unchanged with respect to the transformation
$\nu^{}_\mu \leftrightarrow \nu^{}_\tau$ and thus feature $M^{}_{e \mu} = M^{}_{e \tau}$ and
$M^{}_{\mu \mu} = M^{}_{\tau \tau}$ (with $M^{}_{\alpha \beta}$ for $\alpha, \beta= e, \mu, \tau$
being the matrix elements of $M^{}_\nu$). Such a symmetry (which results in $\theta^{}_{23} = \pi/4$ and
$\theta^{}_{13} =0$) was historically motivated by the experimental facts that $\theta^{}_{23}$
takes a value close to $\pi/4$ while $\theta^{}_{13}$ was only constrained by $\sin^2_{}{2\theta^{}_{13}}
< 0.18 $ \cite{chooz} (and thus might be negligibly small) at the time. However, the relatively large
$\theta^{}_{13} \simeq 0.15$ observed recently \cite{dyb} requires a significant breaking of this symmetry
unless neutrinos are quasi-degenerate in masses \cite{GJP}.
Hence we need to go beyond this simple possibility to accommodate the experimental
results in a better way. In this connection, the $\mu$-$\tau$ reflection symmetry \cite{MTR,review2} may serve as
a unique alternative: When $M^{}_l$ is diagonal, $M^{}_\nu$ should remain invariant under the transformation
\footnote{This operation is a combination of the $\mu$-$\tau$ exchange
and CP conjugate transformations --- a typical kind of the generalized CP transformations \cite{gcp}.}
\begin{eqnarray}
\nu^{}_e \leftrightarrow \nu^c_e \;, \hspace{1cm} \nu^{}_\mu \leftrightarrow \nu^c_\tau \;,
\hspace{1cm} \nu^{}_\tau \leftrightarrow \nu^c_\mu \;,
\label{1.3}
\end{eqnarray}
and thus be characterized by
\begin{eqnarray}
M^{}_{e\mu} = M^*_{e\tau} \;, \hspace{1cm} M^{}_{\mu\mu} = M^*_{\tau\tau}  \;, \hspace{1cm}
M^{}_{ee} = M^*_{ee} \;, \hspace{1cm} M^{}_{\mu\tau} = M^*_{\mu\tau} \;.
\label{1.4}
\end{eqnarray}
In addition to allowing for an arbitrary $\theta^{}_{13}$, this symmetry predicts $\theta^{}_{23} = \pi/4$
and $\delta = \pm \pi/2$ \cite{GL} which are close to the present data, thereby having been
attracting a lot of interests \cite{MTRs}. Moreover, $\rho$ and $\sigma$ are required to
take the trivial values 0 or $\pi/2$.

Nevertheless, it is hard to believe that the $\mu$-$\tau$ reflection symmetry can remain as an
exact one. On the experimental side, the aforementioned results seem to hint towards
$\theta^{}_{23} \neq \pi/4$ (and possibly $\delta \neq \pm \pi/2$). On the theoretical side,
flavor symmetries are generally implemented at a superhigh energy scale and so the renormalization
group (RG) running effect may provide a source for the symmetry breaking as we will see. In view of these
considerations, it is worthwhile to consider the breaking of this symmetry. In the next section,
we perform a systematic study of the possible symmetry-breaking patterns and their implications
for the mixing parameters. First of all, we establish an equation set relating the symmetry-breaking
parameters in an $M^{}_\nu$ of approximate $\mu$-$\tau$ reflection symmetry and the deviations of mixing parameters
from their special values taken in the symmetry context. While the numerical
results for these equations are analyzed in section 2.1, some analytical approximations will
be derived in section 2.2 to explain the corresponding numerical results. In section 2.3 the
general treatment is applied to some specific symmetry breaking arising from the RG running
effect. Finally, we summarize our main results in section 3.

\section{Breaking of the $\mu$-$\tau$ reflection symmetry}

Above all, let us define some parameters to characterize the breaking of $\mu$-$\tau$ reflection
symmetry. For this purpose, one can introduce the parameters
\begin{eqnarray}
\epsilon^{}_{1}=\displaystyle \frac{M^{}_{e \mu}- M^{*}_{e\tau}}{M^{}_{e \mu}+M^{*}_{e\tau}} \;, \hspace{1cm}
\epsilon^{}_{2}=\displaystyle \frac{M^{}_{\mu\mu}-M^{*}_{\tau \tau}}{M^{}_{\mu \mu}+ M^{*}_{\tau\tau}} \;, \hspace{1cm}
\epsilon^{}_{3}=\displaystyle \frac{{\rm Im}(M^{}_{ee})}{{\rm Re}(M^{}_{ee})} \;, \hspace{1cm}
\epsilon^{}_{4}=\displaystyle \frac{{\rm Im}(M^{}_{\mu \tau})}{{\rm Re}(M^{}_{\mu \tau})} \;,
\label{2.1}
\end{eqnarray}
by following the discussions about the breaking of $\mu$-$\tau$ permutation symmetry in Ref. \cite{GJKLST}.
Note that they correspond to the four symmetry conditions in Eq. (\ref{1.4}) one by one.
These parameters have to be small in magnitude (say $|\epsilon^{}_i| \leq 0.1$ for $i=1,2,3,4$)
in order to keep the $\mu$-$\tau$ reflection symmetry as an approximate one. In terms of them,
the most general neutrino mass matrix of an approximate $\mu$-$\tau$ reflection symmetry can always
be parameterized in a manner as follows: Suppose, at the symmetry level, there is a neutrino mass
matrix of the form
\begin{eqnarray}
M^{(0)}_\nu = \left( \begin{matrix}
A^{}_0 &\hspace{0.3cm} B^{}_0 &\hspace{0.3cm} B^*_0\cr
B^{}_0 &\hspace{0.3cm} C^{}_0 &\hspace{0.3cm} D^{}_0 \cr
B^{*}_0&\hspace{0.3cm} D^{}_0 &\hspace{0.3cm} C^{*}_0
\end{matrix} \right) \;,
\label{2.2}
\end{eqnarray}
in which $A^{}_0$ and $D^{}_0$ are real. This neutrino mass matrix can be diagonalized by
a unitary matrix $U^{(0)}$ (an analogue of $U$) with its parameters satisfying the requirements
\begin{eqnarray}
\phi^{(0)}_1= \phi^{(0)}_2 + \phi^{(0)}_3= 0 \;, \hspace{1cm} \theta^{(0)}_{23} = \pi/4 \;,
\hspace{1cm} \delta^{(0)} = \pm \pi/2 \;, \hspace{1cm} \rho^{(0)}, \sigma^{(0)} = 0 \;
\; {\rm or} \; \; \pi/2 \;.
\label{2.3}
\end{eqnarray}
After the symmetry is softly broken, $M^{(0)}_\nu$ may receive
a general perturbation as given by
\begin{eqnarray}
M^{(1)}_\nu = \left( \begin{matrix}
\delta^{}_{ee} & \hspace{0.3cm} \delta^{}_{e\mu} & \hspace{0.3cm} \delta^{}_{e\tau} \cr
\delta^{}_{e\mu} & \hspace{0.3cm} \delta^{}_{\mu\mu} & \hspace{0.3cm} \delta^{}_{\mu\tau} \cr
\delta^{}_{e\tau} & \hspace{0.3cm} \delta^{}_{\mu\tau} & \hspace{0.3cm} \delta^{}_{\tau\tau}
\end{matrix} \right) \;,
\label{2.4}
\end{eqnarray}
which can be decomposed into two parts as
\begin{eqnarray}
M^{(1)}_\nu  =  \frac{1}{2} \left( \begin{matrix}
2{\rm Re}(\delta^{}_{ee}) & \hspace{0.3cm} \delta^{}_{e\mu} + \delta^{*}_{e\tau}
& \hspace{0.3cm} \delta^{*}_{e\mu} + \delta^{}_{e\tau} \cr
\delta^{}_{e\mu} + \delta^{*}_{e\tau} & \hspace{0.3cm}
\delta^{}_{\mu\mu} + \delta^{*}_{\tau\tau} & \hspace{0.3cm} 2{\rm Re}(\delta^{}_{\mu\tau})\cr
\delta^{*}_{e\mu} + \delta^{}_{e\tau} & \hspace{0.3cm} 2{\rm Re}(\delta^{}_{\mu\tau})
& \hspace{0.3cm} \delta^{*}_{\mu\mu}+\delta^{}_{\tau\tau}
\end{matrix} \right)
 +  \frac{1}{2} \left( \begin{matrix}
2{\rm i Im}(\delta^{}_{ee}) & \hspace{0.3cm} \delta^{}_{e\mu} - \delta^{*}_{e\tau}
& \hspace{0.3cm} \delta^{}_{e\tau} - \delta^{*}_{e\mu} \cr
\delta^{}_{e\mu} - \delta^{*}_{e\tau} & \hspace{0.3cm}
\delta^{}_{\mu\mu} - \delta^{*}_{\tau\tau} & \hspace{0.3cm} 2{\rm i Im}(\delta^{}_{\mu\tau}) \cr
\delta^{}_{e\tau} - \delta^{*}_{e\mu} & \hspace{0.3cm} 2{\rm i Im}(\delta^{}_{\mu\tau})
& \hspace{0.3cm} \delta^{}_{\tau\tau} - \delta^{*}_{\mu\mu}
\end{matrix} \right) \;.
\label{2.5}
\end{eqnarray}
Consequently, the complete neutrino mass matrix $M^{}_\nu = M^{(0)}_\nu + M^{(1)}_\nu$
can be parameterized as
\begin{eqnarray}
M^{}_\nu = \left( \begin{matrix}
A (1+{\rm i}\epsilon^{}_3) & \hspace{0.3cm} B(1+\epsilon^{}_1)
& \hspace{0.3cm} B^{*}(1-\epsilon^{*}_1)\cr
B(1+\epsilon^{}_1) & \hspace{0.3cm} C(1+\epsilon^{}_2)
& \hspace{0.3cm}  D(1+{\rm i} \epsilon^{}_4) \cr
B^{*}(1-\epsilon^{*}_1) & \hspace{0.3cm} D(1+{\rm i}\epsilon^{}_4)
& \hspace{0.3cm} C^{*}(1-\epsilon^{*}_2)
\end{matrix} \right) \;,
\label{2.6}
\end{eqnarray}
with
\begin{eqnarray}
& A = A^{}_0 + {\rm Re}(\delta^{}_{ee}) \;, \hspace{0.8cm}
& B = B^{}_0 + \displaystyle \frac{\delta^{}_{e\mu}+\delta^{*}_{e\tau}}{2} \;, \nonumber \\
& D = D^{}_0 + {\rm Re}(\delta^{}_{\mu\tau}) \;, \hspace{0.8cm}
& C = C^{}_0 + \displaystyle \frac{\delta^{}_{\mu\mu}+\delta^{*}_{\tau\tau}}{2} \;,
\label{2.7}
\end{eqnarray}
and
\begin{eqnarray}
\epsilon^{}_1 = \frac{\delta^{}_{e\mu}-\delta^{*}_{e\tau}}{2B} \;, \hspace{1cm}
\epsilon^{}_2 = \frac{\delta^{}_{\mu\mu}-\delta^{*}_{\tau\tau}}{2C} \;, \hspace{1cm}
\epsilon^{}_3 = \frac{{\rm Im}(\delta^{}_{ee})}{A} \;, \hspace{1cm}
\epsilon^{}_4 = \frac{{\rm Im}(\delta^{}_{\mu\tau})}{D} \;.
\label{2.8}
\end{eqnarray}
It should be noted that ${\rm Im}(\epsilon^{}_{1,2})$ and $\epsilon^{}_{3,4}$ will transform in a way as
\begin{eqnarray}
& & {\rm Im}(\epsilon^{}_1) \to {\rm Im}(\epsilon^{}_1) + \varphi^{}_1 + (\varphi^{}_2 + \varphi^{}_3)/2 \;,
\hspace{1cm} \epsilon^{}_3 \to \epsilon^{}_3 + 2 \varphi^{}_1 \;, \nonumber \\
&&  {\rm Im}(\epsilon^{}_2) \to {\rm Im}(\epsilon^{}_2) + \varphi^{}_2 + \varphi^{}_3 \;,
\hspace{1cm} \epsilon^{}_4 \to \epsilon^{}_4 + \varphi^{}_2 + \varphi^{}_3 \;,
\label{2.9}
\end{eqnarray}
under the neutrino-field rephasing
\begin{eqnarray}
& & \nu^{}_{e} \to e^{{\rm i}\varphi^{}_1} \nu^{}_{e}
\simeq (1+ {\rm i} \varphi^{}_1) \nu^{}_e \;, \hspace{1cm}
\nu^{}_{\mu} \to e^{{\rm i}\varphi^{}_2} \nu^{}_{\mu}
\simeq (1+ {\rm i} \varphi^{}_2) \nu^{}_\mu \;, \nonumber \\
& & \nu^{}_{\tau} \to e^{{\rm i}\varphi^{}_3} \nu^{}_{\tau}
\simeq (1+ {\rm i} \varphi^{}_3) \nu^{}_\tau \;,
\label{2.10}
\end{eqnarray}
with $\varphi^{}_{1, 2, 3}$ being some small parameters comparable to $\epsilon^{}_i$.
Taking advantage of such a freedom, one can always achieve $\epsilon^{}_{3, 4}= 0$
from the general case given by Eq. (\ref{2.6}). In the following discussions,
we therefore concentrate on this particular case without loss of generality.

Starting from an $M^{}_\nu$ of the form in Eq. (\ref{2.6}) but with $\epsilon^{}_{3, 4}=0$,
we study dependence of the mixing parameters on $\epsilon^{}_{1, 2}$.
To this end, we diagonalize such an $M^{}_\nu$ with one unitary matrix
in a straightforward way
\begin{eqnarray}
U^\dagger M^{}_\nu U^* = {\rm Diag} (m^{}_1, m^{}_2, m^{}_3) \;.
\label{2.11}
\end{eqnarray}
The mixing parameters in $U$ are expected to lie around those special values
in Eq. (\ref{2.3}) and the corresponding deviations
\begin{eqnarray}
\Delta \phi^{}_1 = \phi^{}_1- 0 \;,  & \hspace{1cm}
\Delta \phi = (\phi^{}_2 + \phi^{}_3)/2 - 0 \;, &  \hspace{1cm}
\Delta \theta = \theta^{}_{23} - \pi/4 \;, \nonumber \\
\Delta \delta = \delta - \delta^{(0)} \;, & \hspace{-0.3cm}
\Delta \rho = \rho - \rho^{(0)} \;, & \hspace{1cm}
\Delta \sigma = \sigma - \sigma^{(0)} \;,
\label{2.12}
\end{eqnarray}
are some small quantities. By making perturbation expansions for these small quantities
in Eq. (\ref{2.11}), one reaches the following relations that connect the mixing-parameter
deviations with $\epsilon^{}_{1,2}$
\begin{eqnarray}
&& m^{}_3 s^2_{13} \Delta \delta + \overline m^{}_1 c^2_{12} \Delta \rho + \overline m^{}_2 s^2_{12} \Delta \sigma
= ( m^{}_3 s^2_{13}-m^{}_{11} ) \Delta \phi^{}_1 \;, \nonumber\\
&& 2 m^{}_{12} \bar s^{}_{13} \Delta \theta- m^{}_{11} s^2_{13} \Delta \delta-\overline m^{}_1 s^2_{12} \Delta \rho
- \overline m^{}_2 c^2_{12} \Delta \sigma =(m^{}_{22}-m^{}_3) \Delta \phi  \;, \nonumber\\
&& [ m^{}_{12} + {\rm i} (m^{}_{11} + m^{}_3) \bar s^{}_{13} ] \Delta \theta
- (m^{}_{11} - m^{}_3) \bar s^{}_{13} \Delta \delta - 2 \overline m^{}_1 c^{}_{12}
( {\rm i} s^{}_{12} + c^{}_{12} \bar s^{}_{13}) \Delta \rho
\nonumber \\
&&  \hspace{0.7cm} + 2 \overline m^{}_{2} s^{}_{12}  ({\rm i} c^{}_{12} - s^{}_{12} \bar s^{}_{13}) \Delta \sigma
= [m^{}_{12} - {\rm i} (m^{}_{11} + m^{}_3) \bar s^{}_{13}]
({\rm i} \Delta \phi^{}_1 + {\rm i} \Delta \phi - \epsilon^{}_1) \;,
\nonumber \\
&&¡¡2 (m^{}_{22} - m^{}_3 ) \Delta \theta - 2(m^{}_{12}-{\rm i} m^{}_{11}\bar s^{}_{13}) \bar s^{}_{13} \Delta \delta
- 2\overline m^{}_1 s^{}_{12} ({\rm i}s^{}_{12} + 2c^{}_{12} \bar s^{}_{13} )\Delta \rho
\nonumber \\
&& \hspace{0.7cm} - 2 \overline m^{}_2 c^{}_{12} ({\rm i}c^{}_{12} - 2s^{}_{12} \bar s^{}_{13} )\Delta \sigma
= (m^{}_{22} + m^{}_3 - 2{\rm i} m^{}_{12} \bar s^{}_{13}) (2{\rm i}\Delta \phi -\epsilon^{}_2) \;.
\label{2.13}
\end{eqnarray}
In order to make the expressions compact, the definitions
\begin{eqnarray}
& & m^{}_{11} = \overline m^{}_1 c^2_{12} + \overline m^{}_2 s^2_{12} \;,  \hspace{1cm}
m^{}_{12} = (\overline m^{}_1 - \overline m^{}_2) c^{}_{12} s^{}_{12} \;,  \hspace{1cm}
m^{}_{22} = \overline m^{}_1 s^2_{12} + \overline m^{}_2 c^2_{12} \;, \nonumber \\
& & \overline m^{}_1 = m^{}_1 {\rm exp}[2 {\rm i} \rho^{(0)}] \;,  \hspace{1.45cm}
\overline m^{}_2 = m^{}_2 {\rm exp}[2 {\rm i} \sigma^{(0)}] \;, \hspace{1.65cm}
\bar s^{}_{13} = -{\rm i} s^{}_{13} {\rm exp}[{\rm i} \delta^{(0)}] \;,
\label{2.14}
\end{eqnarray}
have been taken.

After solving Eq. (\ref{2.13}) we obtain $\Delta \theta$, $\Delta \delta$, $\Delta \rho$
and $\Delta \sigma$ as some linear functions of ${\rm R}^{}_{1,2} = {\rm Re}(\epsilon^{}_{1,2})$ and
${\rm I}^{}_{1,2} = {\rm Im}(\epsilon^{}_{1,2})$, which can be parameterized as
\begin{eqnarray}
& \Delta \theta = c^{\theta}_{r1}{\rm R}^{}_1+c^{\theta}_{i1}{\rm I}^{}_1
+c^{\theta}_{r2}{\rm R}^{}_2+c^{\theta}_{i2}{\rm I}^{}_2 \;, \hspace{1cm}
& \Delta \delta = c^{\delta}_{r1}{\rm R}^{}_1+c^{\delta}_{i1}{\rm I}^{}_1
+c^{\delta}_{r2}{\rm R}^{}_2+c^{\delta}_{i2}{\rm I}^{}_2 \;, \nonumber \\
& \Delta \rho = c^{\rho}_{r1}{\rm R}^{}_1+c^{\rho}_{i1}{\rm I}^{}_1
+c^{\rho}_{r2}{\rm R}^{}_2+c^{\rho}_{i2}{\rm I}^{}_2 \;, \hspace{1cm}
& \Delta \sigma = c^{\sigma}_{r1}{\rm R}^{}_1+c^{\sigma}_{i1}{\rm I}^{}_1
+c^{\sigma}_{r2}{\rm R}^{}_2+c^{\sigma}_{i2}{\rm I}^{}_2 \;.
\label{2.15}
\end{eqnarray}
The coefficients in these expressions measure the sensitive strengths of mixing-parameter
deviations to the symmetry-breaking parameters. For example, $c^{\theta}_{r1}$ measures the
sensitive strength of $\Delta \theta$ to ${\rm R}^{}_1$. The contribution of any given
${\rm R}^{}_1$ to $\Delta \theta$ is expressed as the product of it with
$c^{\theta}_{r1}$ (i.e., $c^{\theta}_{r1} {\rm R}^{}_1$).
There are two things to be noted: For one thing, such a ${\rm R}^{}_1$ will also
contribute to $\Delta \delta$, $\Delta \rho$ and $\Delta \sigma$ by an amount
of $c^{\delta}_{r1} {\rm R}^{}_1$, $c^{\rho}_{r1} {\rm R}^{}_1$ and
$c^{\sigma}_{r1} {\rm R}^{}_1$, respectively.
For another thing, $\Delta \theta$ would receive an additional contribution of $c^{\theta}_{i1}
{\rm I}^{}_1$ ($c^{\theta}_{r2} {\rm R}^{}_2$ or $c^{\theta}_{i2} {\rm I}^{}_2$)
if ${\rm I}^{}_1$ (${\rm R}^{}_2$ or ${\rm I}^{}_2$) were non-vanishing in the meanwhile.
In consideration of the pre-requisition $|\epsilon^{}_{1, 2}| \leq 0.1$, the coefficients
must have magnitudes $\geq \mathcal O(1)$ in order to cause some sizeable
(say $0.1 \simeq 6^\circ$) mixing-parameter deviations. If one coefficient
is much greater than 1 (say 10 or 100), even a tiny (at least 0.01 or 0.001)
symmetry-breaking parameter can give rise to some sizable mixing-parameter deviation.
But if one coefficient is much smaller than 1 (say 0.1 or 0.01), the resulting
mixing-parameter deviation will be negligibly small (at most 0.01 or 0.001).
Although the mixing-parameter deviations are of direct interest, we will first concentrate
on the coefficients and then turn to their implications for the mixing-parameter deviations
for the following considerations: (1) Given any specific symmetry-breaking
pattern (i.e., definite ${\rm R}^{}_{1, 2}$ and ${\rm I}^{}_{1, 2}$) in some physical context
(e.g., the RG-induced symmetry breaking as will be discussed in section 2.3), the resulting
mixing-parameter deviations can be read directly by making use of Eq. (\ref{2.15}) provided
that the coefficients are known. (2) When the mixing-parameters are determined experimentally
to a good degree of accuracy, the required symmetry-breaking pattern may be inferred
with the help of Eq. (\ref{2.15}) provided that the coefficients are known. As one will see,
the values of the coefficients (equivalently the mixing-parameter deviations) are strongly
correlated with the neutrino mass spectrum and the values of $\rho^{(0)}$ and $\sigma^{(0)}$
once the symmetry-breaking strengths (i.e., the values of ${\rm R}^{}_{1, 2}$ and
${\rm I}^{}_{1, 2}$) are specified. In the following, this kind of correlations will be studied
in some detail both numerically and analytically.

\subsection{Numerical results}

In this section, the coefficients are explored in a numerical way.
In Figs. (1-4) we have presented the coefficients (associated with ${\rm R}^{}_1$,
${\rm I}^{}_1$, ${\rm R}^{}_2$ and ${\rm I}^{}_2$ successively) against the lightest
neutrino mass ($m^{}_1$ for NH or $m^{}_3$ for IH) for various combinations of $\rho^{(0)}$
and $\sigma^{(0)}$ (i.e., $[\rho^{(0)}, \sigma^{(0)}] = [0, 0]$, $[\pi/2, 0]$, $[0, \pi/2]$
or $[\pi/2, \pi/2]$). The black, red, green and blue colors are assigned to the coefficients
for $\Delta \theta$, $\Delta \delta$, $\Delta \rho$ and $\Delta \sigma$, respectively.
In order to save space, the absolute value of a coefficient will be shown in
the dashed line if it is negative. By contrast, the full line will be used when the coefficients
are positive. As no observable mixing-parameter deviation will arise
from a highly suppressed coefficient, the region where the coefficients have magnitudes
smaller than 0.01 is not shown. In doing the calculations we have specified
$\delta^{(0)}=-\pi/2$. When it takes the opposite value $\pi/2$, the coefficients will
either simply stay invariant or just change their signs
\begin{eqnarray}
&& c^{\theta}_{r1} \to c^{\theta}_{r1} \;,  \hspace{1.2cm}  c^{\delta}_{r1} \to -c^{\delta}_{r1} \;,
\hspace{1.0cm} c^{\rho}_{r1} \to -c^{\rho}_{r1} \;,  \hspace{1.0cm} c^{\sigma}_{r1} \to -c^{\sigma}_{r1} \;, \nonumber \\
&& c^{\theta}_{i1} \to -c^{\theta}_{i1} \;, \hspace{1cm} c^{\delta}_{i1} \to c^{\delta}_{i1} \;,  \hspace{1.37cm}
c^{\rho}_{i1} \to c^{\rho}_{i1} \;,  \hspace{1.4cm} c^{\sigma}_{i1} \to c^{\sigma}_{i1} \;, \nonumber \\
&& c^{\theta}_{r2} \to c^{\theta}_{r2} \;, \hspace{1.2cm}  c^{\delta}_{r2} \to -c^{\delta}_{r2} \;,
\hspace{1.0cm} c^{\rho}_{r2} \to -c^{\rho}_{r2} \;,  \hspace{1.0cm} c^{\sigma}_{r2} \to -c^{\sigma}_{r2} \;, \nonumber \\
&& c^{\theta}_{i2} \to -c^{\theta}_{i2} \;,  \hspace{1cm} c^{\delta}_{i2} \to c^{\delta}_{i2} \;, \hspace{1.37cm}
c^{\rho}_{i2} \to c^{\rho}_{i2} \;, \hspace{1.4cm} c^{\sigma}_{i2} \to c^{\sigma}_{i2} \;.
\label{2.1.1}
\end{eqnarray}
The point is that Eq. (\ref{2.13}) is invariant with respect to the transformations
\begin{eqnarray}
&& \Delta \theta \to \Delta \theta  \;, \hspace{1cm} \Delta \delta \to - \Delta \delta  \;, \hspace{1cm}
\Delta \rho \to - \Delta \rho  \;, \hspace{1cm} \Delta \sigma \to - \Delta \sigma  \;, \nonumber \\
&& {\rm R}^{}_1 \to {\rm R}^{}_1  \;, \hspace{1.15cm} {\rm I}^{}_1 \to - {\rm I}^{}_1  \;, \hspace{1.4cm}
{\rm R}^{}_2 \to {\rm R}^{}_2  \;, \hspace{1.45cm} {\rm I}^{}_2 \to - {\rm I}^{}_2  \;,
\label{2.1.2}
\end{eqnarray}
combined with $\bar s^{}_{13} \to -\bar s^{}_{13}$ as well as $\Delta \phi^{}_1 \to - \Delta \phi^{}_1$
and $\Delta \phi \to - \Delta \phi$. For reference, in Tables (1-4) we have listed some representative
values of the coefficients (for $\Delta \theta$, $\Delta \delta$, $\Delta \rho$ and $\Delta \sigma$ successively)
at $m^{}_1 (m^{}_3) = 0.001, 0.01$ and 0.1 eV for the NH (IH) case. Numbers in the square brackets denote
the coefficients' values in the IH case. When a coefficient takes values in the range $-0.01\to 0$ or $0 \to 0.01$,
its values will be reported as 0.00 or $-0.00$.

By virtue of the above numerical results one may draw the following conclusions regarding
the coefficients: (1) For $\rho^{(0)} \neq \sigma^{(0)}$ (or $[\rho^{(0)}, \sigma^{(0)}] = [\pi/2, \pi/2]$),
the coefficients get greatly enhanced (or suppressed) when the neutrino masses are quasi-degenerate
$m^{}_1 \simeq m^{}_2 \simeq m^{}_3$ (e.g., the particular case of $m^{}_1 (m^{}_3) \simeq 0.1$ eV).
(2) When the absolute neutrino mass scale is small (e.g., the particular case of
$m^{}_1 (m^{}_3) \simeq 0.001$ eV) and $\rho^{(0)} \neq \sigma^{(0)}$, most of the coefficients
will have a much greater magnitude in the IH case compared to in the NH case.
(3) $\Delta \theta$ is most sensitive to ${\rm R}^{}_{2}$
while $\Delta \delta$, $\Delta \rho$ and $\Delta \sigma$ to all the symmetry-breaking parameters.
In magnitude, the coefficients for $\Delta \delta$, $\Delta \rho$ and $\Delta \sigma$ (which
can even obtain some magnitudes around 100 when the neutrino masses are quasi-degenerate and
$\rho^{(0)} \neq \sigma^{(0)}$) are generally much greater than those for $\Delta \theta$.
Besides these general features, some specific comments for the coefficients
are given in order:
\begin{enumerate}
\item Among the coefficients for $\Delta \theta$, $|c^{\theta}_{r2}|$ is the most significant one
and takes values of $\mathcal O$(1) in most cases. But it decreases to $\mathcal O$(0.1)
in the case of $m^{}_3 \ll m^{}_1 \simeq m^{}_2$ (e.g., the particular
case of $m^{}_3 \simeq 0.001$ eV) combined with $\rho^{(0)} \neq \sigma^{(0)}$.
$|c^{\theta}_{r1}|$ can also reach $\mathcal O$(1) in the case of IH combined with
$\rho^{(0)} \neq \sigma^{(0)}$. $|c^{\theta}_{i1}|$ and $|c^{\theta}_{i2}|$
are well below $\mathcal O$(0.1), indicating that $\Delta \theta$ is insensitive to ${\rm I}^{}_{1,2}$.
\item As for the coefficients for $\Delta \delta$, $|c^{\delta}_{r1}|$ (except in the case of IH
combined with $\rho^{(0)} = \sigma^{(0)}$), $|c^{\delta}_{i1}|$ and $|c^{\delta}_{i2}|$
generally have values of $\mathcal O$(1) or greater. $|c^{\delta}_{r2}|$ can be
significant only in the case of IH combined with $\rho^{(0)} \neq \sigma^{(0)}$.
\item The coefficients for $\Delta \rho$ obtain magnitudes of $\mathcal O$(1)
or greater in most cases, with the exceptions: $|c^{\rho}_{r1}|$ and $|c^{\rho}_{r2}|$
are substantially suppressed in the case of IH combined with $\rho^{(0)} = \sigma^{(0)}$.
The coefficients for $\Delta \sigma$ almost share the same properties as their counterparts
for $\Delta \rho$ except that their magnitudes are somewhat smaller (as a result of $m^{}_2 > m^{}_1$).
\end{enumerate}

Now that the coefficients are known well, we discuss their implications for
$\Delta \theta$ and $\Delta \delta$ (which are of more practical interests than
$\Delta \rho$ and $\Delta \sigma$ since the Majorana phases cannot be pinned
down in a foreseeable future).
\begin{enumerate}
\item In the case of $m^{}_3 \ll m^{}_1 \simeq m^{}_2$ (or $m^{}_1 \simeq m^{}_2 \simeq m^{}_3$)
combined with $\rho^{(0)} \neq \sigma^{(0)}$, $|{\rm R}^{}_1| \simeq 0.1$ (or $\simeq 0.02$)
is capable of producing $|\Delta \theta| \simeq 0.1$. In other cases, $|{\rm R}^{}_1| \leq 0.1$ is unable to
induce sizable $\Delta \theta$. On the other hand, $|{\rm R}^{}_2| \leq 0.1$ can give rise
to sizable $\Delta \theta$ in most cases: If the neutrino masses are quasi-degenerate,
$|\Delta \theta| \simeq 0.1$ may easily arise (except in the
case of $[\rho^{(0)}, \sigma^{(0)}] = [\pi/2, \pi/2]$); If not, $|{\rm R}^{}_2| \simeq 0.1$
will yield $|\Delta \theta|$ around 0.05 (except in the case of IH combined with
$\rho^{(0)} \neq \sigma^{(0)}$). It should be noted that a positive ${\rm R}^{}_1$ or ${\rm R}^{}_2$
always contributes a positive (or negative) $\Delta \theta$ in the NH (or IH) case.
In comparison, $|{\rm I}^{}_1| \leq 0.1$ and $|{\rm I}^{}_2| \leq 0.1$ cannot lead to sizable
$\Delta \theta$ unless the neutrino masses are quasi-degenerate and $\rho^{(0)} \neq \sigma^{(0)}$.
\item For $m^{}_1 \ll m^{}_2 \ll m^{}_3$ (e.g., the particular case of
$m^{}_1 \simeq 0.001$ eV), $|{\rm R}^{}_1| \simeq 0.1$ leads
$|\Delta \delta|$ close to 0.1. In the case of $m^{}_3 \ll m^{}_1 \simeq m^{}_2$
(or $m^{}_1 \simeq m^{}_2 \simeq m^{}_3$) combined with $\rho^{(0)} \neq \sigma^{(0)}$,
sizable $\Delta \delta$ can arise from $|{\rm R}^{}_1|$ or $|{\rm R}^{}_2|$ as small as
0.01 (or 0.001). In other cases, $|{\rm R}^{}_1| \leq 0.1$ and $|{\rm R}^{}_2| \leq 0.1$
have no chance to generate sizable $\Delta \delta$. In the $m^{}_1 \ll m^{}_2 \ll m^{}_3$
case, $|{\rm I}^{}_1| \simeq 0.1$ brings about $|\Delta \delta|$ around 0.05.
For $m^{}_3 \ll m^{}_1 \simeq m^{}_2$, $|{\rm I}^{}_1|$ as small as
0.01 (or 0.001) may trigger sizable $\Delta \delta$ in the case of
$\rho^{(0)} = \sigma^{(0)}$ (or $\rho^{(0)} \neq \sigma^{(0)}$). When the neutrino
masses are quasi-degenerate, even a tiny ${\rm I}^{}_1$ is able to induce
sizable $\Delta \delta$ (except in the case of
$[\rho^{(0)}, \sigma^{(0)}] = [\pi/2, \pi/2]$).
The contributions of ${\rm I}^{}_2$ to $\Delta \delta$ bear many similarities
with those of ${\rm I}^{}_1$.
\end{enumerate}

For illustration, we give a toy example to show how to make use of the above results.
In this connection, we discuss how the global-fit results $\theta^{}_{23}/^\circ = 41.6^{+1.5}_{-1.2}$
and $\delta/^\circ = 261^{+51}_{-59}$ in the NH case \cite{global2} may arise from an approximate
$\mu$-$\tau$ reflection symmetry. (For simplicity, only the best-fit results will be used.)
Since $\Delta \theta$ is most sensitive to ${\rm R}^{}_2$, one wonders whether
a single ${\rm R}^{}_2$
\footnote{Of course, in a realistic context, the mixing-parameter deviations may receive
contributions from not merely one symmetry-breaking parameter.}
can give rise to appropriate $\Delta \theta$ and $\Delta \delta$ simultaneously.
This requires $c^{\delta}_{r2}/ c^{\theta}_{r2} = \Delta \delta/\Delta \theta = (-9^\circ)/(-3.4^\circ)
\simeq 2.65$. With the aid of Fig. 3, it turns out that the coefficients have chance to
fulfill such a requirement in the case of $[\rho^{(0)}, \sigma^{(0)}] = [\pi/2, 0]$
(see the sub-figure labelled by ``NH $-$+''): At $m^{}_1 = 0.019$ eV, the values of $c^{\theta}_{r2}$,
$c^{\delta}_{r2}$, $c^{\rho}_{r2}$ and $c^{\sigma}_{r2}$ respectively read 0.78, 2.07, $-$2.45 and
$-$0.90. In this case, a ${\rm R}^{}_2 \simeq -0.076$ will result in $\theta^{}_{23} \simeq 41.6^\circ$
and $\delta = 261^\circ$ (as well as $\rho \simeq 100.7^\circ$ and $\sigma \simeq 3.9^\circ$).

Finally, we discuss the consequences of breaking of $\mu$-$\tau$ reflection symmetry on
the allowed range of effective Majorana neutrino mass $|M^{}_{ee}|$ which directly controls
the rates of neutrinoless double-beta decays \cite{0nbb}. For this purpose, one obtains
\begin{eqnarray}
{\rm Re}(M^{}_{ee}) & \simeq & \overline m^{}_1 c^2_{12} c^2_{13} + \overline m^{}_2 s^2_{12} c^2_{13}
- m^{}_3 s^2_{13} \;, \nonumber \\
{\rm Im}(M^{}_{ee}) & \simeq &  \overline m^{}_1 c^2_{12} \Delta \rho + \overline m^{}_2 s^2_{12} \Delta \sigma
+m^{}_3 s^2_{13} \Delta \delta + ( m^{}_{11} - m^{}_3 s^2_{13} ) \Delta \phi^{}_1 \;.
\end{eqnarray}
Because the symmetry-breaking parameter $\epsilon^{}_3 = {\rm Im}(M^{}_{ee})/{\rm Re}(M^{}_{ee})$
should be a small quantity (e.g., $|\epsilon^{}_3| \le 0.1$) if we want to maintain
the $\mu$-$\tau$ reflection symmetry as an approximate one, the value of
\begin{eqnarray}
|M^{}_{ee}| = \sqrt{[{\rm Re}(M^{}_{ee})]^2 + [{\rm Im}(M^{}_{ee})]^2}
\simeq | {\rm Re}(M^{}_{ee})| \left(1+ \frac{\epsilon^2_3}{2} \right) \;,
\end{eqnarray}
can be well approximated by that of $|{\rm Re}(M^{}_{ee})|$. It is thus fair to say
that the consequences of breaking of $\mu$-$\tau$ reflection symmetry on the allowed range
of $|M^{}_{ee}|$ are negligibly small. In Fig. 5 we present the possible values of
$|{\rm Re}(M^{}_{ee})|$ as a function of the lightest neutrino mass
$m^{}_1$ (or $m^{}_3$) in the NH (or IH) case for various combinations
of $\rho^{(0)}$ and $\sigma^{(0)}$ \cite{Mohapatra}. (1) In the NH case, the three components
of $|{\rm Re}(M^{}_{ee})|$ add constructively to a maximal level for $[\rho^{(0)}, \sigma^{(0)}]
= [\pi/2, \pi/2]$. By contrast, the three components will cancel each other out (i.e.,
$|{\rm Re}(M^{}_{ee})| \simeq 0$) at $m^{}_1 \simeq 0.002$ eV (or 0.007 eV) for
$[\rho^{(0)}, \sigma^{(0)}] = [\pi/2, 0]$ (or $[0, \pi/2]$). (2) In the IH case,
the value of $|{\rm Re}(M^{}_{ee})|$ is mainly determined by the first two
components as the third one is highly suppressed. Because of $m^{}_1
\simeq m^{}_2$ in the IH case, $|{\rm Re}(M^{}_{ee})|$ approximates
to $m^{}_1 $ (or $m^{}_1 (c^2_{12} - s^2_{12})$) for $\rho^{(0)} = \sigma^{(0)}$
(or $\rho^{(0)} \neq \sigma^{(0)}$).

\begin{figure}[h]
\centering
\includegraphics[width=6.5in]{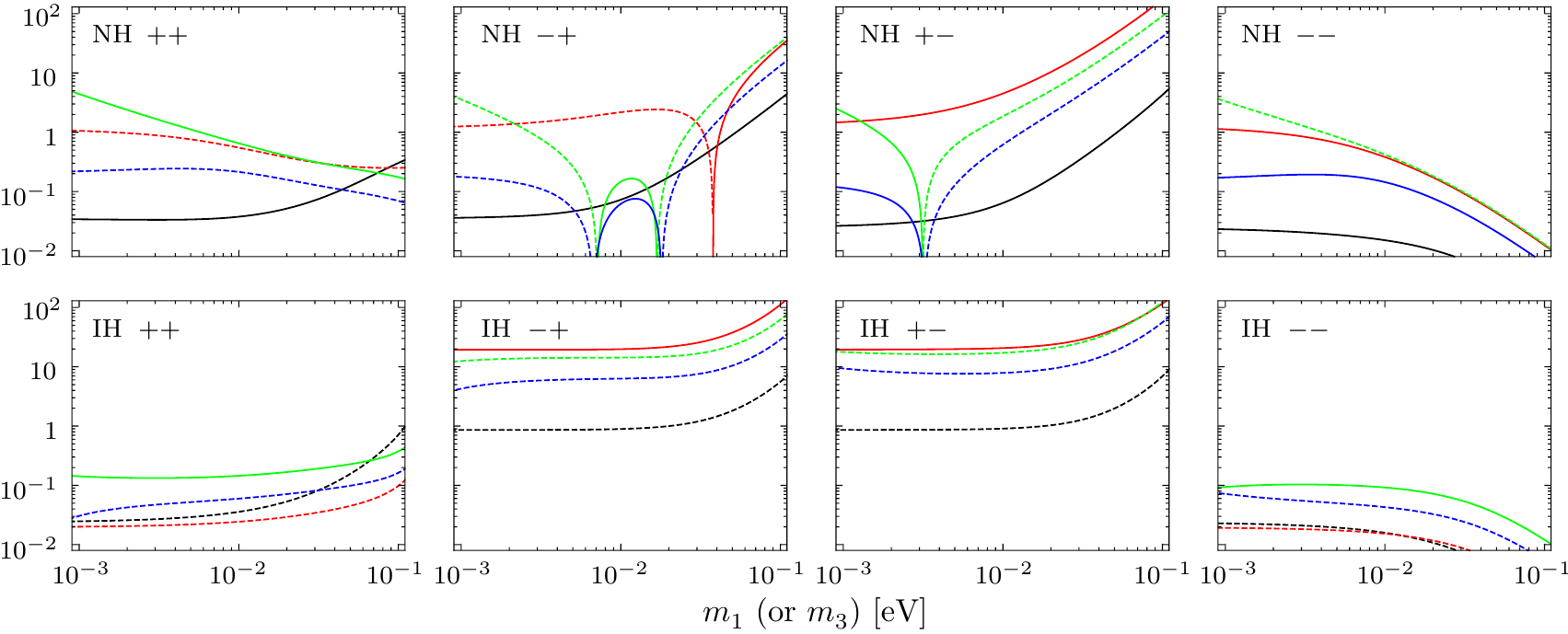}
\caption{The coefficients associated with ${\rm R^{}_1}$ ($c^{\theta}_{r1}$ in black, $c^{\delta}_{r1}$ in red,
$c^{\rho}_{r1}$ in green and $c^{\sigma}_{r1}$ in blue) against the lightest neutrino mass
$m^{}_1$ (or $m^{}_3$) in the NH (or IH) case for various combinations
of $\rho^{(0)}$ and $\sigma^{(0)}$ with $\delta^{(0)}=-\pi/2$. The signs $++, -+, +-$ and $++$
respectively stand for $[\rho^{(0)}, \sigma^{(0)}] = [0, 0], [\pi/2, 0], [0, \pi/2]$ and $[\pi/2, \pi/2]$.}
\end{figure}

\begin{figure}[h]
\centering
\includegraphics[width=6.5in]{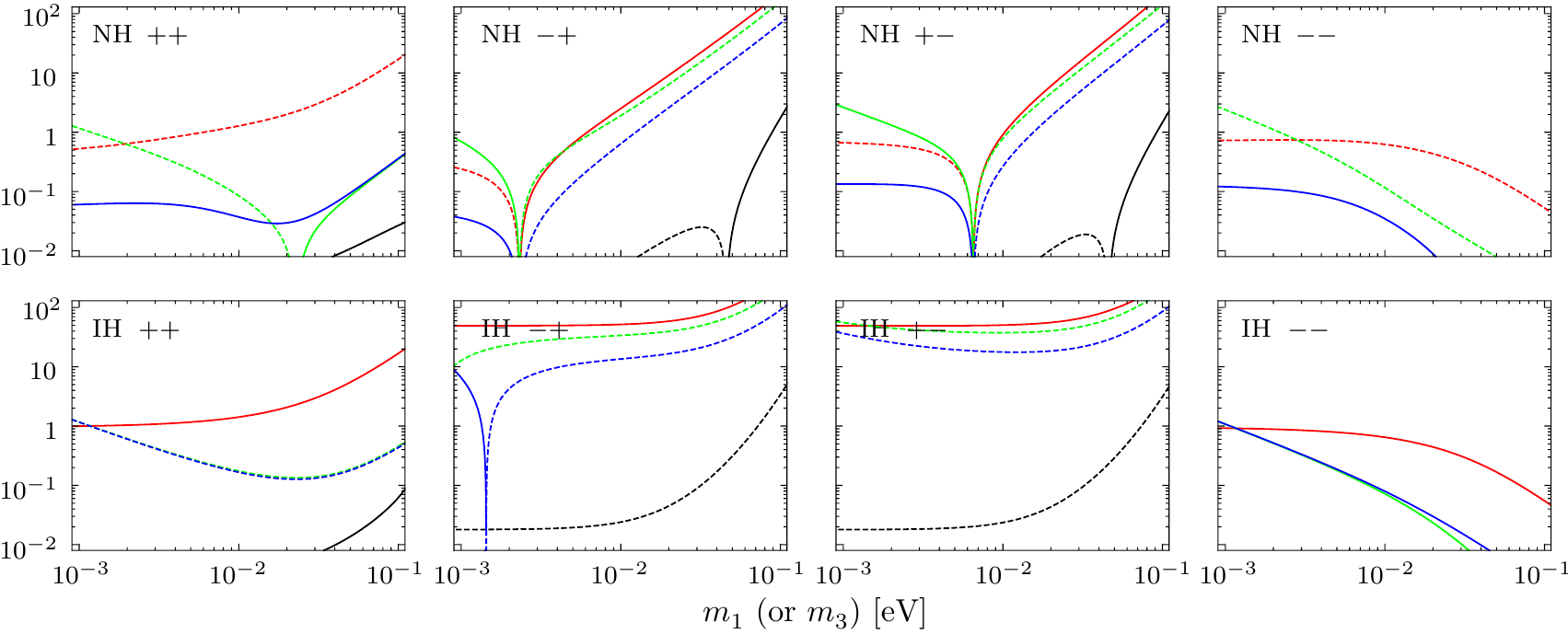}
\caption{The coefficients associated with ${\rm I^{}_1}$ ($c^{\theta}_{i1}$ in black, $c^{\delta}_{i1}$ in red,
$c^{\rho}_{i1}$ in green and $c^{\sigma}_{i1}$ in blue) against the lightest neutrino mass
$m^{}_1$ (or $m^{}_3$) in the NH (or IH) case for various combinations
of $\rho^{(0)}$ and $\sigma^{(0)}$ with $\delta^{(0)}=-\pi/2$. The signs $++, -+, +-$ and $++$
respectively stand for $[\rho^{(0)}, \sigma^{(0)}] = [0, 0], [\pi/2, 0], [0, \pi/2]$ and $[\pi/2, \pi/2]$.}
\end{figure}

\begin{figure}
\centering
\includegraphics[width=6.5in]{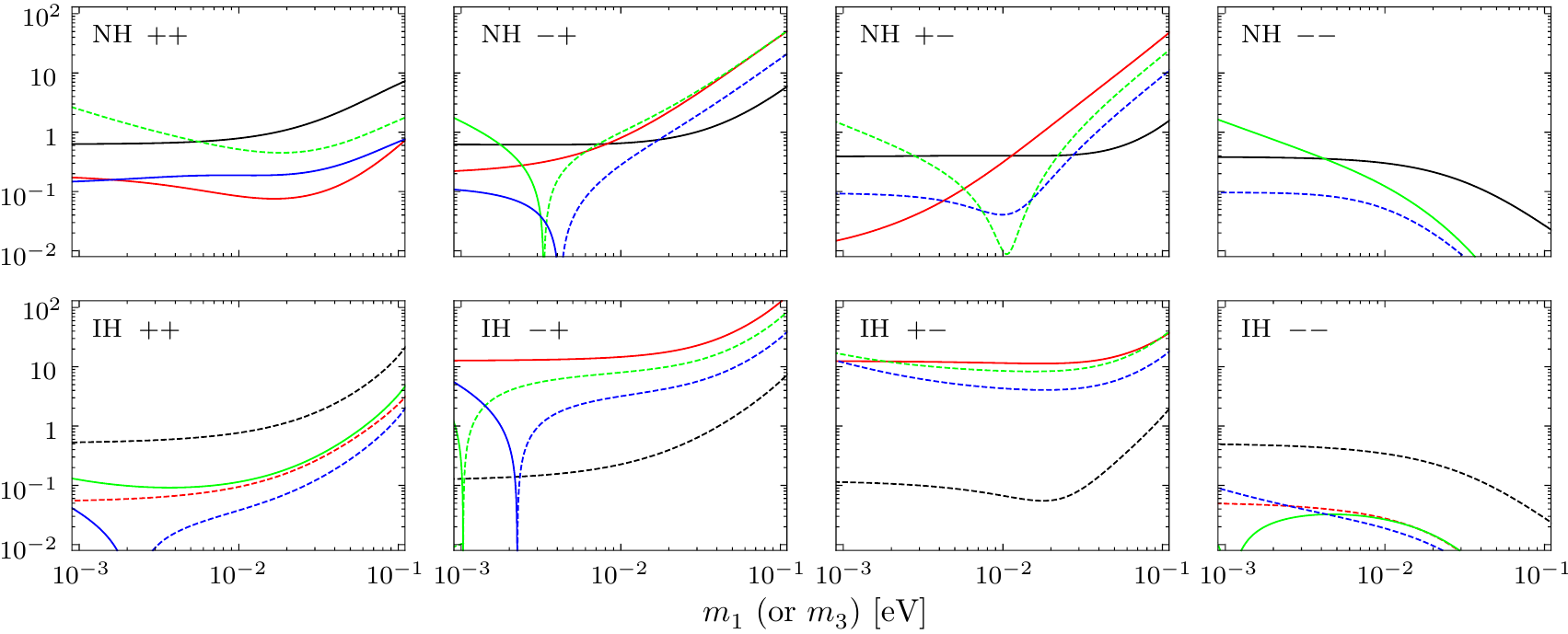}
\caption{The coefficients associated with ${\rm R^{}_2}$ ($c^{\theta}_{r2}$ in black, $c^{\delta}_{r2}$ in red,
$c^{\rho}_{r2}$ in green and $c^{\sigma}_{r2}$ in blue) against the lightest neutrino mass
$m^{}_1$ (or $m^{}_3$) in the NH (or IH) case for various combinations
of $\rho^{(0)}$ and $\sigma^{(0)}$ with $\delta^{(0)}=-\pi/2$. The signs $++, -+, +-$ and $++$
respectively stand for $[\rho^{(0)}, \sigma^{(0)}] = [0, 0], [\pi/2, 0], [0, \pi/2]$ and $[\pi/2, \pi/2]$.}
\end{figure}

\begin{figure}
\centering
\includegraphics[width=6.5in]{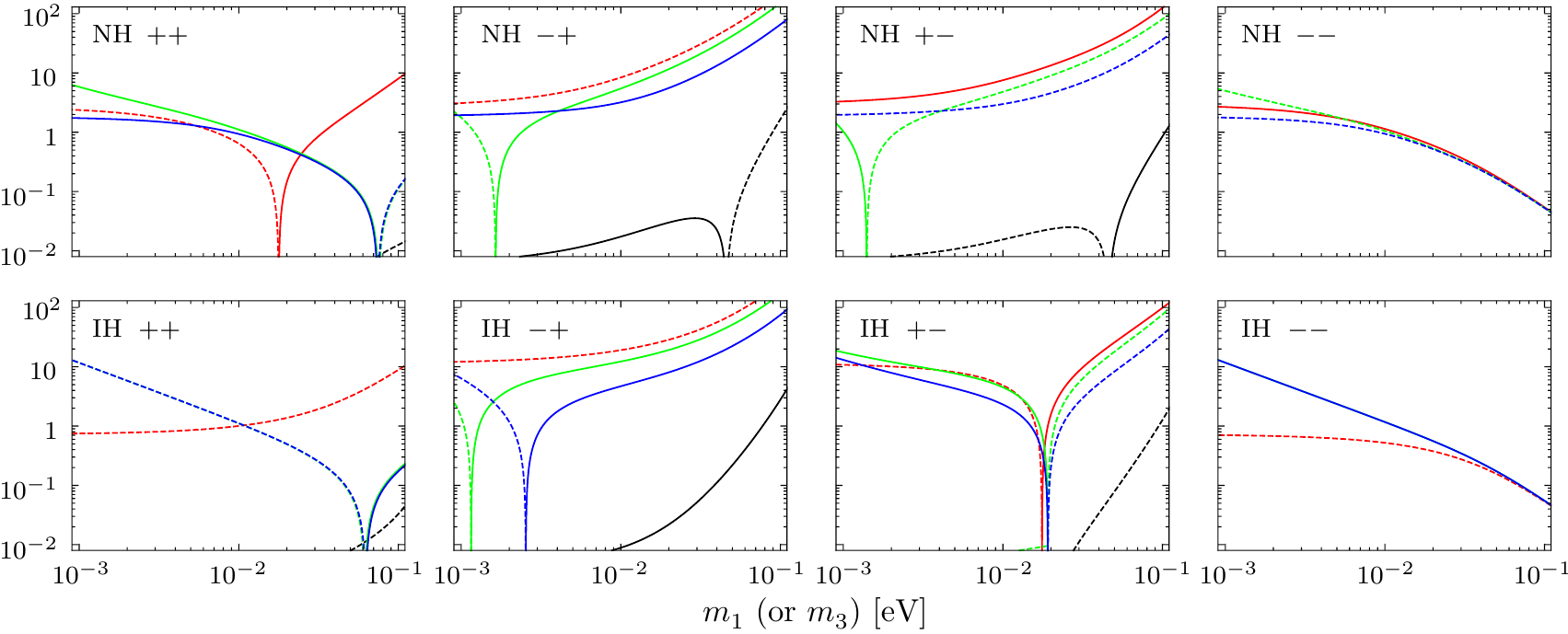}
\caption{The coefficients associated with ${\rm I^{}_2}$ ($c^{\theta}_{i2}$ in black, $c^{\delta}_{i2}$ in red,
$c^{\rho}_{i2}$ in green and $c^{\sigma}_{i2}$ in blue) against the lightest neutrino mass
$m^{}_1$ (or $m^{}_3$) in the NH (or IH) case for various combinations
of $\rho^{(0)}$ and $\sigma^{(0)}$ with $\delta^{(0)}=-\pi/2$. The signs $++, -+, +-$ and $++$
respectively stand for $[\rho^{(0)}, \sigma^{(0)}] = [0, 0], [\pi/2, 0], [0, \pi/2]$ and $[\pi/2, \pi/2]$.}
\end{figure}

\begin{table}[h]
\centering
\begin{tabular}{|p{3.5cm}<{\centering}|p{1.cm}<{\centering}|p{3.cm}<{\centering}|p{3.cm}<{\centering}|p{3.cm}<{\centering}|} \hline
\backslashbox{[$\rho^{(0)}$, $\sigma^{(0)}$]}{$m^{}_1 [m^{}_3]$} &  & 0.001 eV & $0.01$ eV &  $0.1$ eV \\ \hline \hline
$[0, 0]$  & $c^{\theta}_{r1}$ & 0.03 [$-$0.02]   & 0.04 [$-$0.04] & 0.30 [$-$0.71]  \\ \hline
$[\pi/2, 0]$ &  & 0.04 [$-$0.86] & 0.07 [$-$0.89] & 3.64 [$-$5.59]  \\ \hline
$[0, \pi/2]$ & & 0.03 [$-$0.86]  &  0.06 [$-$0.91] & 4.27 [$-$6.95]  \\ \hline
$[\pi/2, \pi/2]$ & & 0.02 [$-$0.02] & 0.02 [$-$0.02]  & 0.00 [$-$0.00] \\ \hline \hline
$[0, 0]$   & $c^{\theta}_{i1}$ & 0.00 [0.00]   & 0.00 [0.00] & 0.03 [0.06]  \\ \hline
$[\pi/2, 0]$ & & 0.00 [$-$0.02] & $-$0.01 [$-$0.02] & 1.74 [$-$3.38]  \\ \hline
$[0, \pi/2]$ & & 0.00 [$-$0.02]  & $-$0.00 [$-$0.02] & 1.47 [$-$3.13]  \\ \hline
$[\pi/2, \pi/2]$ & & 0.00 [0.00] & 0.00 [0.00]  & 0.00 [0.00] \\ \hline \hline
$[0, 0]$  & $c^{\theta}_{r2}$ & 0.63 [$-$0.53] & 0.79 [$-$0.77] & 6.48 [$-$15.3]  \\ \hline
$[\pi/2, 0]$ & & 0.62 [$-$0.13] & 0.65 [$-$0.23] & 4.93 [$-$5.72]  \\ \hline
$[0, \pi/2]$ & & 0.39 [$-$0.11]  & 0.40 [$-$0.07] & 1.30 [$-$1.44]  \\ \hline
$[\pi/2, \pi/2]$ & & 0.38 [$-$0.49] & 0.30 [$-$0.34]  & 0.03 [$-$0.03] \\ \hline \hline
$[0, 0]$   & $c^{\theta}_{i2}$ & 0.01 [$-$0.00] & 0.00 [$-$0.00] & $-$0.01 [$-$0.03]  \\ \hline
$[\pi/2, 0]$ & & 0.01 [0.00] & 0.02 [0.01] & $-$1.66 [2.78]  \\ \hline
$[0, \pi/2]$ & & $-$0.01 [0.00]  & $-$0.02 [0.00] & 0.85 [$-$1.27]  \\ \hline
$[\pi/2, \pi/2]$ & & $-$0.01 [$-$0.00] &$-$0.00 [$-$0.00]  & $-$0.00 [$-$0.00]  \\ \hline
\end{tabular}
\caption{Some representative values of the coefficients for $\Delta \theta$
( $c^{\theta}_{r1}$, $c^{\theta}_{i1}$, $c^{\theta}_{r2}$ and $c^{\theta}_{i2}$ successively)
at $m^{}_1 (m^{}_3) = 0.001, 0.01$ and 0.1 eV in the NH (IH) case
for various combinations of $\rho^{(0)}$ and $\sigma^{(0)}$ with $\delta^{(0)}=-\pi/2$.}
\end{table}

\begin{table}[h]
\centering
\begin{tabular}{|p{3.5cm}<{\centering}|p{1cm}<{\centering}|p{3.5cm}<{\centering}|p{3.cm}<{\centering}|p{3.cm}<{\centering}|} \hline
\backslashbox{[$\rho^{(0)}$, $\sigma^{(0)}$]}{$m^{}_1 [m^{}_3]$} & & 0.001 eV & 0.01 eV &  0.1 eV \\ \hline \hline
$[0, 0]$  & $c^{\delta}_{r1}$ & $-$1.06 [$-$0.02] & $-$0.55 [$-$0.02] & $-$0.25 [$-$0.10]  \\ \hline
$[\pi/2, 0]$ & & $-$1.25 [19.4] & $-$2.18 [19.8] & 28.3 [112]  \\ \hline
$[0, \pi/2]$ & & 1.48 [19.5] & 4.50 [20.7] & 173 [117]  \\ \hline
$[\pi/2, \pi/2]$ & & 1.12 [$-$0.02] & 0.38 [$-$0.02]  & 0.01 [$-$0.00]  \\ \hline \hline
$[0, 0]$  & $c^{\delta}_{i1}$ & $-$0.53 [1.00] & $-$1.28 [1.42] & $-$17.0 [16.8]  \\ \hline
$[\pi/2, 0]$  & & $-$0.24[49.2] & 2.52 [51.8] & 220 [315] \\ \hline
$[0, \pi/2]$  & & $-$0.66 [49.0] & 0.91 [50.5] & 216 [243]  \\ \hline
$[\pi/2, \pi/2]$ & & $-$0.73 [0.92] & $-$0.62 [0.65]  & $-$0.05 [0.05]  \\ \hline \hline
$[0, 0]$  & $c^{\delta}_{r2}$ & 0.17 [$-$0.06] & 0.08 [$-$0.09] & 0.58 [$-$2.34]  \\ \hline
$[\pi/2, 0]$  & & 0.22[12.7] & 0.80 [14.7] & 42.0 [124] \\ \hline
$[0, \pi/2]$  & & 0.02 [12.4] & 0.31 [11.6] & 38.6 [31.6]  \\ \hline
$[\pi/2, \pi/2]$ & & 0.00 [$-$0.05] & $-$0.00 [$-$0.03]  & 0.00 [$-$0.00]  \\ \hline \hline
$[0, 0]$  & $c^{\delta}_{i2}$ & $-$2.37 [$-$0.75] & $-$0.64 [$-$1.00] & 8.01 [$-$8.88]  \\ \hline
$[\pi/2, 0]$ & & $-$3.10[$-$12.2] & $-$8.39 [$-$19.1] & $-$211 [$-$259] \\ \hline
$[0, \pi/2]$ & & 3.31 [$-$10.8] & 7.54 [$-$4.87] & 125 [98.6]  \\ \hline
$[\pi/2, \pi/2]$ & & 2.66 [$-$0.70] & 1.14 [$-$0.52]  & 0.05 [$-$0.05]  \\ \hline
\end{tabular}
\caption{Some representative values of the coefficients for $\Delta \delta$ ($c^{\delta}_{r1}$, $c^{\delta}_{i1}$,
$c^{\delta}_{r2}$ and $c^{\delta}_{i2}$ successively) at $m^{}_1 (m^{}_3) = 0.001, 0.01$ and 0.1 eV in the NH (IH) case
for various combinations of $\rho^{(0)}$ and $\sigma^{(0)}$ with $\delta^{(0)}=-\pi/2$.}
\end{table}

\begin{table}[h]
\centering
\begin{tabular}{|p{3.5cm}<{\centering}|p{1cm}<{\centering}|p{3.5cm}<{\centering}|p{3.cm}<{\centering}|p{3.cm}<{\centering}|} \hline
\backslashbox{[$\rho^{(0)}$, $\sigma^{(0)}$]}{$m^{}_1 [m^{}_3]$} & & 0.001 eV & 0.01 eV &  0.1 eV \\ \hline \hline
$[0, 0]$  & $c^{\rho}_{r1}$ & 4.42 [0.14] & 0.65 [0.14] & 0.17 [0.37]  \\ \hline
$[\pi/2, 0]$ & & $-$3.57 [$-$12.5] & 0.15 [$-$14.3] & $-$32.4 [$-$62.5] \\ \hline
$[0, \pi/2]$ & & 2.19 [$-$17.7] & $-$1.85 [$-$17.2] & $-$91.4 [$-$124] \\ \hline
$[\pi/2, \pi/2]$ & & $-$3.32 [0.10] & $-$0.42 [0.09]  & $-$0.01 [0.01]  \\ \hline \hline
$[0, 0]$ & $c^{\rho}_{i1}$ & $-$1.17 [$-$1.17] & $-$0.08 [$-$0.17] & 0.35 [$-$0.45]  \\ \hline
$[\pi/2, 0]$ & & 0.70 [$-$12.7] & $-$1.91 [$-$33.8] & $-$157 [$-$207]  \\ \hline
$[0, \pi/2]$ & & 2.63 [$-$55.7] & $-$0.79 [$-$37.5] & $-$141 [$-$190] \\ \hline
$[\pi/2, \pi/2]$ & & $-$2.39 [1.07] & $-$0.12 [0.07]  & $-$0.00 [0.00]  \\ \hline \hline
$[0, 0]$  & $c^{\rho}_{r2}$ & $-$2.43 [0.12] & $-$0.52 [0.11] & $-$1.58 [3.37]  \\ \hline
$[\pi/2, 0]$ & & 1.52 [0.23] & $-$0.99 [$-$8.02] & $-$42.7 [$-$67.0]  \\ \hline
$[0, \pi/2]$  & & $-$1.34 [$-$16.1] & $-$0.01 [$-$8.51] & $-$19.6 [$-$32.1] \\ \hline
$[\pi/2, \pi/2]$ & & 1.47 [$-$0.00] & 0.12 [0.03]  & 0.00 [0.00]  \\ \hline \hline
$[0, 0]$  & $c^{\rho}_{i2}$ & 5.82 [$-$11.7] & 1.11 [$-$1.11] & $-$0.11 [0.18]  \\ \hline
$[\pi/2, 0]$ & & $-$1.76 [$-$1.40] & 5.46 [12.3] & 150 [170] \\ \hline
$[0, \pi/2]$ & & 1.03 [17.5] & $-$4.80 [4.51] & $-$80.7 [$-$76.2] \\ \hline
$[\pi/2, \pi/2]$ & & $-$5.00 [11.7] & $-$1.05 [1.17]  & $-$0.05 [0.05]  \\ \hline
\end{tabular}
\caption{Some representative values of  the coefficients for $\Delta \rho$ ($c^{\rho}_{r1}$, $c^{\rho}_{i1}$,
$c^{\rho}_{r2}$ and $c^{\rho}_{i2}$ successively) at $m^{}_1 (m^{}_3) = 0.001, 0.01$ and 0.1 eV in the NH (IH) case
for various combinations of $\rho^{(0)}$ and $\sigma^{(0)}$ with $\delta^{(0)}=-\pi/2$.}
\end{table}

\begin{table}[h]
\centering
\begin{tabular}{|p{3.5cm}<{\centering}|p{1cm}<{\centering}|p{3.5cm}<{\centering}|p{3.cm}<{\centering}|p{3cm}<{\centering}|} \hline
\backslashbox{[$\rho^{(0)}$, $\sigma^{(0)}$]}{$m^{}_1 [m^{}_3]$} & & 0.001 eV & 0.01 eV & 0.1 eV \\ \hline \hline
$[0, 0]$  & $c^{\sigma}_{r1}$ & $-$0.22 [$-$0.03] & $-$0.21 [$-$0.06] & $-$0.07 [$-$0.16]  \\ \hline
$[\pi/2, 0]$ & & $-$0.18 [$-$4.27] & 0.06 [$-$6.31] & $-$13.3 [$-$28.8] \\ \hline
$[0, \pi/2]$ & & 0.12 [$-$9.35] & $-$0.62 [$-$7.88] & $-$40.7 [$-$56.3] \\ \hline
$[\pi/2, \pi/2]$ & & 0.17 [$-$0.07] & 0.14 [$-$0.04]  & 0.01 [$-$0.01] \\ \hline \hline
$[0, 0]$  & $c^{\sigma}_{i1}$ & 0.06 [$-$1.16] & 0.04 [$-$0.17] & 0.36 [$-$0.43]  \\ \hline
$[\pi/2, 0]$ & & 0.04 [6.52] & $-$0.64 [$-$13.5] & $-$68.3 [$-$90.9] \\ \hline
$[0, \pi/2]$ & & 0.13 [$-$36.5] & $-$0.26 [$-$17.7] & $-$63.2 [$-$85.8] \\ \hline
$[\pi/2, \pi/2]$ & & 0.12 [1.08] & 0.03 [0.08]  & $-$0.00 [0.00] \\ \hline \hline
$[0, 0]$  & $c^{\sigma}_{r2}$ & 0.15 [0.03] & 0.19 [$-$0.04] & 0.69 [$-$1.42]  \\ \hline
$[\pi/2, 0]$ & & 0.11 [4.61] & $-$0.28 [$-$3.19] & $-$17.4 [$-$30.8]  \\ \hline
$[0, \pi/2]$ & & $-$0.09 [$-$11.8] & $-$0.04 [$-$4.32] & $-$8.82 [$-$14.7] \\ \hline
$[\pi/2, \pi/2]$ & & $-$0.10 [$-$0.08] & $-$0.05 [$-$0.02]  & $-$0.00 [$-$0.00] \\ \hline \hline
$[0, 0]$  & $c^{\sigma}_{i2}$ & 1.73 [$-$11.7] & 0.93 [$-$1.11] & $-$0.12 [0.17]  \\ \hline
$[\pi/2, 0]$  & & 1.94 [$-$6.36] & 3.19 [4.69] & 65.7 [75.0] \\ \hline
$[0, \pi/2]$  & & $-$1.97 [13.1] & $-$2.99 [2.31] & $-$36.6 [$-$34.8] \\ \hline
$[\pi/2, \pi/2]$ & & $-$1.76 [11.7] & $-$0.94 [1.17]  & $-$0.05 [0.05]  \\ \hline
\end{tabular}
\caption{Some representative values of the coefficients for $\Delta \sigma$ ($c^{\sigma}_{r1}$, $c^{\sigma}_{i1}$,
$c^{\sigma}_{r2}$ and $c^{\sigma}_{i2}$ successively) at $m^{}_1 (m^{}_3) = 0.001, 0.01$ and 0.1 eV in the NH (IH) case
for various combinations of $\rho^{(0)}$ and $\sigma^{(0)}$ with $\delta^{(0)}=-\pi/2$.}
\end{table}

\begin{figure}[h]
\begin{minipage}[t]{0.49\textwidth}
\includegraphics[width=2.8in]{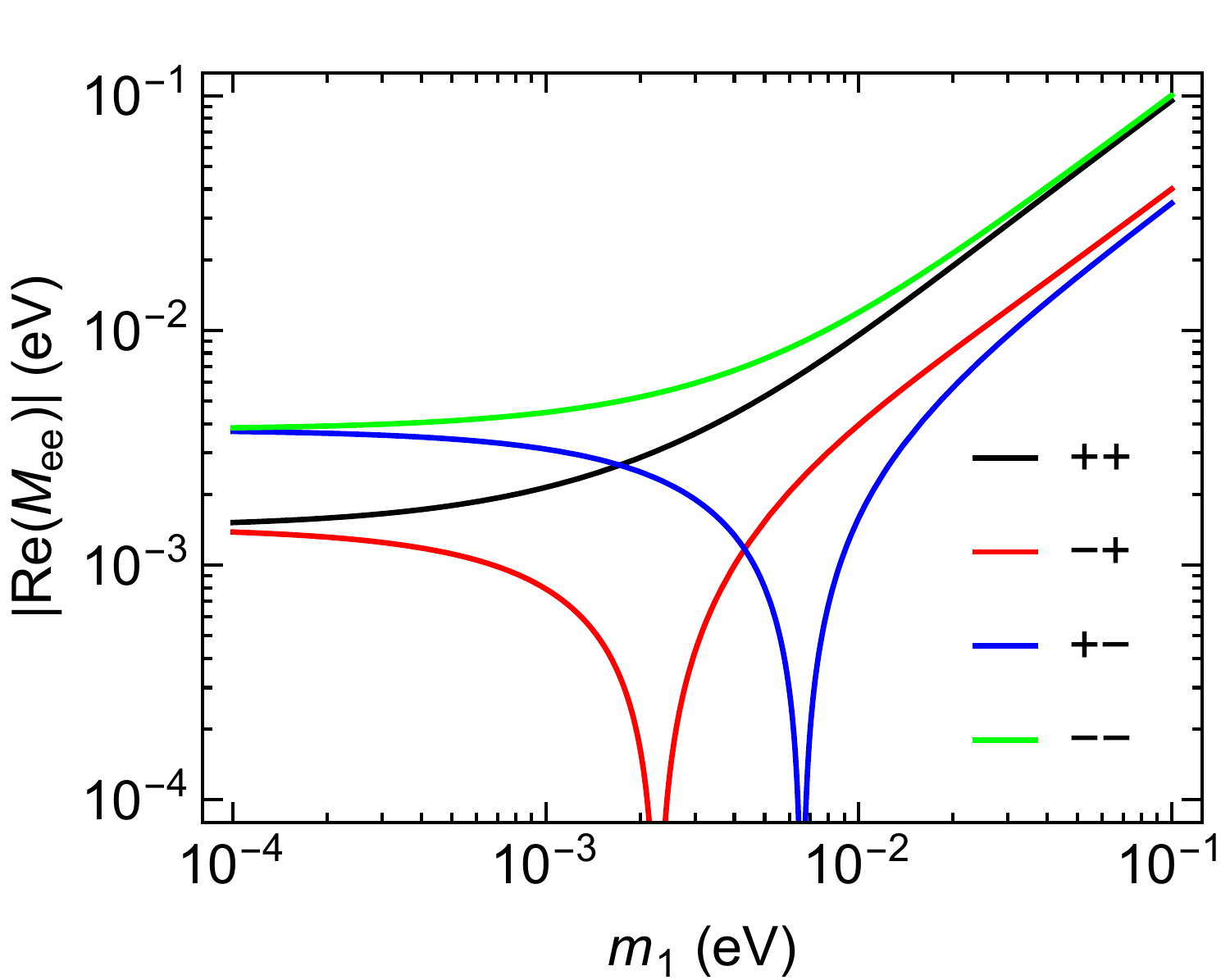}
\end{minipage}
\begin{minipage}[t]{0.49\textwidth}
\includegraphics[width=2.8in]{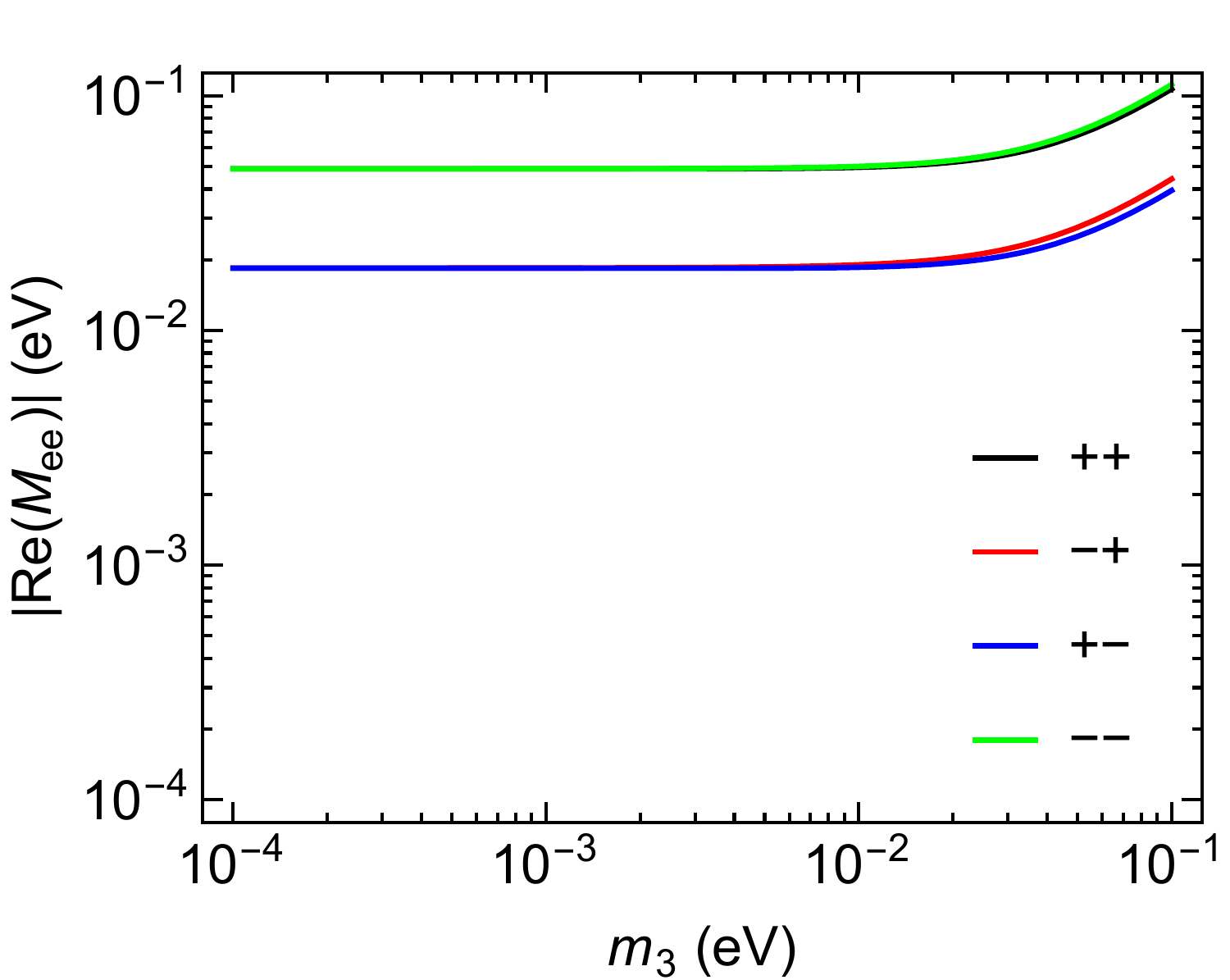}
\end{minipage}
\caption{The possible values of $|{\rm Re}(M^{}_{ee})|$ against the lightest neutrino mass
$m^{}_1$ (or $m^{}_3$) in the NH (or IH) case for various combinations
of $\rho^{(0)}$ and $\sigma^{(0)}$. The signs $++, -+, +-$ and $++$ respectively stand for
$[\rho^{(0)}, \sigma^{(0)}] = [0, 0], [\pi/2, 0], [0, \pi/2]$ and $[\pi/2, \pi/2]$.}
\end{figure}

\subsection{Analytical approximations}

In this section, we give the analytical expressions of $\Delta \delta$ and $\Delta \theta$
to explain the numerical results.
After a lengthy but straightforward calculation, one obtains the approximation results
\begin{eqnarray}
\Delta \delta & \simeq &
[2 {\rm T} (\overline m^{}_1 + \overline m^{}_2) (\overline m^{}_1 - m^{}_3)
(\overline m^{}_2 - m^{}_3) \bar s^{}_{13} ]^{-1}
\{ [2 {\rm T} m^{}_{12} (m^{}_{22} - m^{}_3) (\overline m^{}_1 + \overline m^{}_2- 2 m^{}_{3} s^2_{13}) \nonumber \\
&&  -4m^2_{12} (m^{}_{11} + m^{}_3 + {\rm T} m^{}_{12} ) s^2_{13}
+ 4(\overline m^{}_1 - m^{}_3)(\overline m^{}_2 - m^{}_3) (m^{}_{11} + m^{}_3) s^2_{13}]{\rm R}^{}_1 -[4 m^{}_{12}  \nonumber \\
&& \times(\overline m^{}_1 - m^{}_3)(\overline m^{}_2 - m^{}_3) -2 {\rm T} (\overline m^{}_1 + \overline m^{}_2
- 2 m^{}_3 s^2_{13}) (m^{}_{11} + m^{}_3) (m^{}_{22} -m^{}_3) - 4 {\rm T} \Omega  \nonumber \\
&&  \times m^2_{12} m^{}_3 s^2_{13} ] \bar s^{}_{13} {\rm I}^{}_1 -[ {\rm T} m^{}_{12} (m^{}_{22}
+ m^{}_3) (\overline m^{}_1 + \overline m^{}_2 - 2 m^{}_{3} s^2_{13}) -2 {\rm T} m^{}_{12} (\overline m^{}_1 - m^{}_3)  \nonumber \\
&& \times  ( \overline m^{}_2- m^{}_3) s^2_{13} -2 (m^{}_{11} - m^{}_3)(m^{}_{22}
+ m^{}_3)(m^{}_{11} + m^{}_3 + {\rm T} m^{}_{12} ) s^2_{13}] {\rm R}^{}_2  \nonumber \\
&& + {\rm T} (m^{}_{22} + m^{}_3) [m^2_{12}- (m^{}_{22} - m^{}_3)(2m^{}_{11} + m^{}_{22} - m^{}_3) ]
\bar s^{}_{13} {\rm I}^{}_2 \}  \;, \nonumber \\
\Delta \theta & \simeq &
[2{\rm T} (\overline m^{}_1 + \overline m^{}_2) (\overline m^{}_1 - m^{}_3)(\overline m^{}_2 - m^{}_3)]^{-1}
\{ 2{\rm T} (\overline m^{}_1 + \overline m^{}_2) [m^2_{12} - (m^{2}_{11}-m^{2}_3) s^2_{13}]{\rm R}^{}_1 \nonumber \\
& & +4 m^{}_{11} m^{}_{12} [ {\rm T} (\overline m^{}_1 + \overline m^{}_2)+ 2  m^{}_{12} s^2_{13}] \bar s^{}_{13} {\rm I}^{}_1
-{\rm T} (\overline m^{}_1 + \overline m^{}_2) (m^{}_{11} - m^{}_3)(m^{}_{22} + m^{}_3){\rm R}^{}_2 \nonumber \\
& & -{\rm T} m^{}_{12}[ (\overline m^{}_1 + \overline m^{}_2)  (2m^{}_{11} + m^{}_{22} - m^{}_3)
- 2 m^{}_{11} (m^{}_{22}+ m^{}_3) s^2_{13} ]\bar s^{}_{13} {\rm I}^{}_2 \} \;,
\label{2.2.1}
\end{eqnarray}
with ${\rm T}=\tan{2\theta^{}_{12}}$ and $\Omega = (m^{}_{11} + m^{}_3)/(m^{}_{11} - m^{}_3)$.

For illustration, we discuss the possible values of $\Delta \delta$ and $\Delta \theta$ in several
typical cases. (1) In the case of $m^{}_1 \ll m^{}_2 \simeq \sqrt{\Delta m^2_{21}} \ll m^{}_3
\simeq \sqrt{|\Delta m^2_{31}|}$,
$\Delta \delta$ and $\Delta \theta$ approximate to
\begin{eqnarray}
\Delta \delta & \simeq & \frac{m^{}_3} {2 \overline m^{}_2  \bar s^{}_{13}}
\left[\left( 2 r c^{}_{12} s^{}_{12} + \frac{4}{{\rm T} } s^2_{13} \right) {\rm R}^{}_1 -\bar s^{}_{13} {\rm I}^{}_2
+\left(r c^{}_{12} s^{}_{12} - \frac{2}{{\rm T} } s^2_{13} \right) {\rm R}^{}_2 \right]  - 2s^{2}_{12} {\rm I}^{}_1 \;, \nonumber \\
\Delta \theta & \simeq & s^2_{13} {\rm R}^{}_1 -2 |r| c^{}_{12} s^3_{12} \bar s^{}_{13} {\rm I}^{}_1
+ \frac{1}{2} {\rm R}^{}_2 - \frac{\overline m^{}_{2}} {2m^{}_3} c^{}_{12} s^{}_{12} \bar s^{}_{13} {\rm I}^{}_2 \;,
\label{2.2.2}
\end{eqnarray}
where $r = \Delta m^2_{21}/\Delta m^2_{31}$ has a value of $\pm 0.03$ for NH or IH.
One can see that $\Delta \delta$ is susceptible to the symmetry breaking with the relevant
coefficients having magnitudes of $\mathcal O(1)$, while $\Delta \theta$ is only sensitive to ${\rm R}^{}_2$.
(2) For $m^{}_1 \simeq m^{}_2 \simeq \sqrt{|\Delta m^2_{31}|} \gg m^{}_3$,
in which case one has $m^{}_2 - m^{}_1 \simeq
\Delta m^2_{21}/(2\sqrt{|\Delta m^2_{31}|})$,
the results are strongly dependent on the combinations of $\rho^{(0)}$ and $\sigma^{(0)}$.
If they take the same value, $\Delta \delta$ and $\Delta \theta$ are simplified to
\begin{eqnarray}
\Delta \delta & \simeq & {\rm I}^{}_1 - \frac{3}{4}{\rm I}^{}_2 + \frac{1}{4 {\rm T} \bar s^{}_{13}}
\left[ \left(4 s^2_{13} -2 {\rm T} |r|  c^{}_{12} s^{}_{12} \right){\rm R}^{}_1
+ \left( 2s^2_{13}+ {\rm T} |r|  c^{}_{12}s^{}_{12} \right) {\rm R}^{}_2 \right] \;, \nonumber \\
\Delta \theta & \simeq & -s^2_{13}{\rm R}^{}_1 - |r| c^{}_{12}s^{}_{12} \bar s^{}_{13}
\left( {\rm I}^{}_1-\frac{3}{4}{\rm I}^{}_2 \right)-\frac{1}{2}{\rm R}^{}_2  \;.
\label{2.2.3}
\end{eqnarray}
Obviously, $|c^\delta_{i1}|$ and $|c^\delta_{i2}|$ are close to 1 but $|c^\delta_{r1}|$ and $|c^\delta_{r2}|$
are suppressed to a high level. Among the coefficients for $\Delta \theta$, only $|c^\theta_{r2}|$ is sizable.
When $\rho^{(0)}$ and $\sigma^{(0)}$ differ from each other, one will have
\begin{eqnarray}
\Delta \delta & \simeq & \frac{1}{|r|}
\left[-32 c^3_{12} s^3_{12} \bar s^{}_{13} {\rm R}^{}_1+ 4 \cos{2\theta^{}_{12}} {\rm I}^{}_1
- 4 c^{}_{12} s^{}_{12} \bar s^{}_{13} {\rm R}^{}_2 - 4 c^{2}_{12} s^{2}_{12}
\cos{2\theta^{}_{12}} {\rm I}^{}_2 \right]\;, \nonumber \\
\Delta \theta & \simeq & -4c^{2}_{12}s^2_{12}{\rm R}^{}_1
- c^{}_{12} s^{}_{12} \cos{2\theta^{}_{12}} \left[1 - \frac{4}{|r|}
\cos{2\theta^{}_{12}} s^2_{13}\right] \bar s^{}_{13}
\left( 4 {\rm I}^{}_1 - {\rm I}^{}_2 \right) -\frac{1}{2} \cos^2{2\theta^{}_{12}} {\rm R}^{}_2  \;.
\label{2.2.4}
\end{eqnarray}
Enhanced by the factor $1/|r|$, the coefficients for $\delta$ can easily obtain some magnitudes $\geq 10$.
While $|c^\theta_{r1}|$ with a value about 1 turns out to be the greatest coefficient
for $\Delta \theta$.
(3) When it comes to $m^{}_1 \simeq m^{}_2 \simeq m^{}_3 \simeq m^{}_0$,
in which case $m^{}_2 - m^{}_1 \simeq \Delta m^2_{21}/(2 m^{}_0)$ and
$m^{}_3 - m^{}_1 \simeq \Delta m^2_{31}/(2 m^{}_0)$, the coefficients may get remarkably magnified in some cases.
For $[ \rho^{(0)}, \sigma^{(0)}] = [0, 0]$, $\Delta \delta$ and $\Delta \theta$ are approximately given by
\begin{eqnarray}
\Delta \delta & \simeq & \frac{1}{ {\rm T} \bar s^{}_{13} } \left[ \left( {\rm T} r  c^{}_{12} s^{}_{12} +2 s^2_{13} \right) {\rm R}^{}_1
+\frac{2m^2_0}{\Delta m^2_{31}} \left( {\rm T} r c^{}_{12} s^{}_{12} -2 s^2_{13} \right) {\rm R}^{}_2 \right]
- \frac{2m^2_0}{\Delta m^2_{31}} \left( 2{\rm I}^{}_1- {\rm I}^{}_2 \right)  \;,  \nonumber \\
\Delta \theta & \simeq & \frac{2 m^2_0}{\Delta m^2_{31}} \left[ 2s^2_{13} {\rm R}^{}_1
-r c^{}_{12} s^{}_{12} \bar s^{}_{13} \left(2 {\rm I}^{}_1 -{\rm I}^{}_2 \right)+ {\rm R}^{}_2 \right] \;.
\label{2.2.5}
\end{eqnarray}
At $m^{}_0 \simeq 0.1$ eV, $|c^\delta_{i1}|$ and $|c^\delta_{i2}|$ are around 10
while $|c^\delta_{r1}|$ and $|c^\delta_{r2}|$ are only of $\mathcal O(0.1)$.
$|c^{\theta}_{r2}|$ (with a value close to 10) is still the biggest coefficient
for $\Delta \theta$. In the case of $[\rho^{(0)}, \sigma^{(0)}] = [\pi/2, 0]$, $\Delta \delta$
and $\Delta \theta$ appear as
\begin{eqnarray}
\Delta \delta & \simeq & \frac{2m^2_0}{\Delta m^2_{21}} \left\{ \left[ \frac{4}{\rm T} s^2_{12}
- 2c^{}_{12} s^{}_{12} (1+ 2s^2_{12}) \right]\bar s^{}_{13}  {\rm R}^{}_1 + 2 \cos{2\theta^{}_{12}} {\rm I}^{}_1
-2 c^{}_{12} s^{}_{12} \bar s^{}_{13} {\rm R}^{}_2 - c^2_{12} {\rm I}^{}_2 \right\}  \;,  \nonumber \\
\Delta \theta & \simeq & \frac{2m^2_0}{\Delta m^2_{31}} \left[ 2c^2_{12} s^2_{12} {\rm R}^{}_1
-\frac{4m^2_0}{\Delta m^2_{21}} c^{}_{12} s^{}_{12} \cos{2\theta^{}_{12}} \bar s^{3}_{13}
\left( 2 \cos{2\theta^{}_{12}} {\rm I}^{}_1 - c^2_{12} {\rm I}^{}_2 \right)
+ c^4_{12} {\rm R}^{}_2 \right]\;.
\label{2.2.6}
\end{eqnarray}
When $[\rho^{(0)}, \sigma^{(0)}] = [0, \pi/2]$, the results become
\begin{eqnarray}
\Delta \delta & \simeq & \frac{2m^2_0}{\Delta m^2_{21}} \left\{
-\left[\frac{4}{\rm T}c^2_{12}+2 c^{}_{12} s^{}_{12} (1+ 2c^2_{12}) \right]\bar s^{}_{13}  {\rm R}^{}_1
+ 2 \cos{2\theta^{}_{12}} {\rm I}^{}_1
- 2 c^{}_{12} s^{}_{12} \bar s^{}_{13} {\rm R}^{}_2 + s^2_{12} {\rm I}^{}_2 \right\} \;,  \nonumber \\
\Delta \theta & \simeq & \frac{2m^2_0}{\Delta m^2_{31}} \left[ 2c^2_{12} s^2_{12} {\rm R}^{}_1
- \frac{4m^2_0}{\Delta m^2_{21}} c^{}_{12} s^{}_{12} \cos{2\theta^{}_{12}} \bar s^{3}_{13}
\left( 2 \cos{2\theta^{}_{12}} {\rm I}^{}_1 + s^2_{12} {\rm I}^{}_2 \right)
+ s^4_{12} {\rm R}^{}_2 \right] \;.
\label{2.2.7}
\end{eqnarray}
For these two cases, the coefficients for $\Delta \delta$ may easily obtain a magnitude around 100
owing to the enhancement factor $m^2_0/\Delta m^2_{21}$, while those for $\Delta \theta$ just have
some magnitudes of $\mathcal O(1)$ as the factor $m^2_0/\Delta m^2_{31}$ is not so significant.
Finally, $[ \rho^{(0)}, \sigma^{(0)}] = [\pi/2, \pi/2]$ will lead us to
\begin{eqnarray}
\Delta \delta & \simeq & - \frac{\Delta m^2_{31}}{4 m^2_0 \bar s^{}_{13}} \left[ \left( r c^{}_{12} s^{}_{12}
+ \frac{2}{{\rm T}} s^2_{13} \right) {\rm R}^{}_1 + \bar s^{}_{13} \left({\rm I}^{}_1 - {\rm I}^{}_2 \right) \right] \;,  \nonumber \\
\Delta \theta & \simeq & \frac{\Delta m^2_{31}}{8 m^2_0} \left[2 s^2_{13} {\rm R}^{}_1
-2 r c^{}_{12} s^{}_{12} \bar s^{}_{13} \left({\rm I}^{}_1 - {\rm I}^{}_2 \right)+ {\rm R}^{}_2 \right] \;.
\label{2.2.8}
\end{eqnarray}
It is easy to see that all the coefficients are vanishingly small in this case.
One will find that all the above analytical results agree well with the corresponding
numerical results.

\subsection{RG induced symmetry breaking}

This section is devoted to the RG-induced breaking of $\mu$-$\tau$ reflection symmetry.
A flavor symmetry \cite{review} together with the associated new fields is
usually introduced at an energy scale $\Lambda^{}_{\rm FS}$ much higher than
the electroweak (EW) one $\Lambda^{}_{\rm EW}$. In this case one must consider
the RG running effect when confronting the flavor-symmetry model with the
low-energy data \cite{OZ}. During the RG evolution process the significant
difference between $m^{}_{\mu}$ and $m^{}_{\tau}$ can serve as a unique source
for the breaking of $\mu$-$\tau$ reflection symmetry. As a result,
the general symmetry breaking studied in the above finds an interesting
application in such a specific situation \cite{rge,rge2}. The energy dependence of
neutrino mass matrix is described by its RG equation, which at the one-loop level
appears as \cite{RGE}
\begin{eqnarray}
16 \pi^2 \frac{{\rm d}M^{}_{\nu}}{{\rm d}t} = C \left(Y^{\dagger}_l Y^{}_l
\right)^{T} M^{}_\nu + C M^{}_\nu \left(Y^{\dagger}_l Y^{}_l \right) + \alpha M^{}_{\nu} \;.
\label{2.3.1}
\end{eqnarray}
Here $t$ is defined as $\ln (\mu/\mu^{}_0)$ with $\mu$ denoting the renormalization scale,
whereas $C$ and $\alpha$ read
\begin{eqnarray}
&& C=-\frac{3}{2} \;, \hspace{1cm} \alpha \simeq -3g^2_2+6 y^2_t + \lambda \;,
\hspace{1cm} {\rm in \  the \ SM} \;;  \nonumber \\
&& C=1 \;, \hspace{1cm} \alpha \simeq -\frac{6}{5}g^2_1-6 g^2_2+6 y^2_t \;,
\hspace{1cm} {\rm in \  the \ MSSM} \;.
\label{2.3.2}
\end{eqnarray}
In Eq. (\ref{2.3.1}) the $\alpha$-term is flavor universal and therefore
just contributes an overall rescaling factor (which will be referred to as
$I^{}_{\alpha}$), while the other two terms may modify the structure of
$M^{}_\nu$. In the basis under study, the Yukawa coupling matrix of three
charged leptons is given by $Y^{}_l = {\rm Diag}(y^{}_e, y^{}_\mu, y^{}_\tau)$.
In light of $y^{}_{e} \ll y^{}_\mu \ll y^{}_\tau$, it is reasonable to
neglect the contributions of $y^{}_{e}$ and $y^{}_\mu$.
Integration of the RG equation enables us to connect the neutrino mass matrix
$M^{}_\nu(\Lambda^{}_{\rm FS})$ at $\Lambda^{}_{\rm FS}$ with the corresponding one
at $\Lambda^{}_{\rm EW}$ in a manner as \cite{IRGE}
\begin{eqnarray}
M^{}_\nu(\Lambda^{}_{\rm EW})= I^{}_{\alpha} I^{\dagger}_\tau
M^{}_{\nu}(\Lambda^{}_{\rm FS}) I^{*}_{\tau} \;,
\label{2.3.3}
\end{eqnarray}
where $I^{}_\tau \simeq {\rm Diag} \{1, 1, 1-\Delta^{}_{\tau}\}$ and
\begin{eqnarray}
I^{}_{\alpha} = {\rm exp}\left( \frac{1}{16\pi^2} \int^{\ln \Lambda^{}_{\rm
EW}}_{\ln \Lambda^{}_{\rm FS}} \alpha {\rm d}t \right) , \hspace{1cm}
\Delta^{}_{\tau} = \frac{C}{16\pi^2} \int^{\ln \Lambda^{}_{\rm FS}}_{\ln \Lambda^{}_{\rm
EW}}y^2_{\tau} {\rm d}t \;.
\label{2.3.4}
\end{eqnarray}
In the SM case, the RG running effect is negligible due to the smallness of $y^{}_\tau \simeq 0.01$
(which renders $\Delta^{}_\tau \simeq \mathcal O(10^{-5})$ ).
By contrast, $y^{2}_\tau = (1+ \tan^2{\beta}) m^2_\tau/v^2$ can be enhanced by a large $\tan{\beta}$
in the MSSM case. Given $\Lambda^{}_{\rm FS} \simeq 10^{13}$ GeV, for example, the value of $\Delta^{}_{\tau}$
depends on $\tan{\beta}$ in a way as
\begin{eqnarray}
\Delta^{}_{\tau} \simeq 0.042 \left( \displaystyle \frac{\tan{\beta}}{50} \right)^2 \;.
\end{eqnarray}

With the help of Eq. (\ref{2.3.3}), one will get the RG-corrected neutrino mass matrix at $\Lambda^{}_{\rm EW}$
\begin{eqnarray}
M^{}_{\nu}(\Lambda^{}_{\rm EW}) \simeq I^{}_{\alpha} \left[ M^{}_\nu(\Lambda^{}_{\rm FS})
- \Delta^{}_\tau \left( \begin{matrix} 0 & 0 & M^{}_{e \tau} \cr 0 & 0 & M^{}_{\mu \tau} \cr
M^{}_{e \tau} & M^{}_{\mu \tau} & 2 M^{}_{\tau\tau}\cr
\end{matrix} \right) \right] \;,
\label{2.3.5}
\end{eqnarray}
from a neutrino mass matrix respecting the $\mu$-$\tau$ reflection symmetry at $\Lambda^{}_{\rm FS}$.
By means of the above-mentioned treatment, one may arrange
$M^{}_{\nu}(\Lambda^{}_{\rm EW})$ in a form as given by Eq. (\ref{2.6}) with
$\epsilon^{}_2 =2 \epsilon^{}_1 = \Delta^{}_\tau$ and $\epsilon^{}_{3, 4} = 0$,
implying that $\Delta^{}_\tau$ is the only quantity for measuring the symmetry-breaking strength.
The relations between the mixing-parameter deviations and $\Delta^{}_\tau$ can therefore
be obtained by simply taking ${\rm R^{}_2} = 2{\rm R^{}_1} = \Delta^{}_\tau$ and ${\rm I^{}_{1, 2}}
=0$ in Eq. (\ref{2.13}). By solving these equations numerically,
in Fig. 5 we display $c^\theta_\tau = \Delta \theta/\Delta^{}_\tau$,
$c^\delta_\tau = \Delta \delta/\Delta^{}_\tau$,
$c^\rho_\tau = \Delta \rho/\Delta^{}_\tau$ and
$c^\sigma_\tau = \Delta \rho/\Delta^{}_\tau$
against the lightest neutrino mass $m^{}_1$ ($m^{}_3$) in the NH (IH) case
for various combinations of $\rho^{(0)}$ and $\sigma^{(0)}$ with $\delta^{(0)} = - \pi/2$.
In addition, some representative values of them at $m^{}_1 (m^{}_3) = 0.001, 0.01$ and 0.1 eV for the NH (IH)
case in various situations are presented in Table 5. Since the relations $c^\eta_\tau=c^\eta_{r1}/2+ c^\eta_{r2}$
(for $\eta= \theta, \delta, \rho$ and $\sigma$) hold, the coefficients associated with $\Delta^{}_\tau$
closely resemble those associated with $r^{}_{1, 2}$ in a few aspects:
(1) Their magnitudes tend to grow with the absolute neutrino mass scale (except in the case
of $[\rho^{(0)}, \sigma^{(0)}] = [\pi/2, \pi/2]$).
(2) $|c^\theta_\tau|$ generally takes a value of $\mathcal O(1)$ in most cases,
while $|c^\delta_\tau|$, $|c^\rho_\tau|$ and $|c^\sigma_\tau|$ may
easily reach $\mathcal O(10)$ in the case of IH combined with $\rho^{(0)} \neq \sigma^{(0)}$.
(3) $|c^\delta_\tau|$ and $|c^\rho_\tau|$ are comparable
to each other in magnitude, while $|c^\sigma_\tau|$ is somewhat smaller.
(4) $c^\theta_\tau$ is always positive (negative) in the NH (IH) case. With the help of these
results, one can learn how much $\Delta \theta$ and $\Delta \delta$ are allowed by
$\Delta^{}_\tau \leq 0.04$ (for $\tan{\beta} \leq 50$): (1) In the case of $m^{}_1 \ll m^{}_2
\ll m^{}_3$ or $m^{}_3 \ll m^{}_1 \simeq m^{}_2$, $|\Delta \theta|$ is no greater than 0.02.
When the neutrino masses are quasi-degenerate (except in the case of $[\rho^{(0)}, \sigma^{(0)}]
=[\pi/2, \pi/2]$), $\Delta^{}_\tau \simeq 0.02$ (corresponding to $\tan{\beta} \simeq 35$)
can lead to $|\Delta \theta| \simeq 0.1$.
(2) $|\Delta \delta|$ is also no greater than 0.02 for $m^{}_1 \ll m^{}_2 \ll m^{}_3$.
When $m^{}_3 \ll m^{}_1 \simeq m^{}_2$ is concerned, $|\Delta \delta|$ may
reach 0.1 from $\Delta^{}_\tau \simeq 0.005$ (or will be negligibly small) for
$\rho^{(0)} \neq \sigma^{(0)}$ (or $\rho^{(0)} = \sigma^{(0)}$). In the case of
$m^{}_1 \simeq m^{}_2 \simeq m^{}_3$ combined with $\rho^{(0)} \neq \sigma^{(0)}$,
even $\Delta^{}_\tau \simeq 0.001$ can give rise to sizable $\Delta \delta$.

On the other hand, the analytical expressions for $c^\theta_\tau$, $c^\delta_\tau$
and $c^\rho_\tau$ are found to be
\begin{eqnarray}
c^\theta_\tau &\simeq&
[2(\overline m^{}_1 - m^{}_3)(\overline m^{}_2 - m^{}_3)]^{-1} [m^2_{12} -(m^{}_{11} - m^{}_3)(m^{}_{22} + m^{}_3)]  \;,
\nonumber \\
c^\delta_\tau &\simeq& - \left[ {\rm T} (\overline m^{}_1 + \overline m^{}_2)
(\overline m^{}_1 - m^{}_3) (\overline m^{}_2 - m^{}_3) \overline s^{}_{13} \right]^{-1}
[ {\rm T} m^{}_{12} m^{}_3 (\overline m^{}_1 + \overline m^{}_2- 2 m^{}_3 s^2_{13})  \nonumber \\
& & - 2 (m^{}_{11}+ {\rm T} m^{}_{12}+m^{}_3 ) (\overline m^{}_1  \overline m^{}_2 - m^{}_{22} m^{}_3 ) s^2_{13} ]  \;,
\nonumber \\
c^\rho_\tau & \simeq &- [ 2 \overline m^{}_1 m^{}_3 (\overline m^{}_1 + \overline m^{}_2) (\overline m^{}_1 - m^{}_3)
(\overline m^{}_2 - m^{}_3) \bar s^{}_{13} t^{}_{12} ]^{-1}
\{ {\rm T} m^{}_{12} m^{2}_3 (\overline m^{}_1 + \overline m^{}_2) ( m^{}_{11} \nonumber \\
&& - m^{}_3 s^2_{13} + 2 m^{}_{22} t^2_{12} s^2_{13})  - m^{}_{11} [ {\rm T} m^{}_{12} m^{}_3
(\overline m^{}_1 + \overline m^{}_2- 2 m^{}_3 s^2_{13}) - 2 (m^{}_{11}+ {\rm T} m^{}_{12} \nonumber \\
&& +m^{}_3 ) (\overline m^{}_1  \overline m^{}_2 - m^{}_{22} m^{}_3 ) s^2_{13} ]
[ m^{}_3 + m^{}_{11} \Omega s^2_{13} - m^{}_{22} t^2_{12} s^2_{13}] \} \;,
\label{2.3.6}
\end{eqnarray}
with $t^{}_{12} = \tan{\theta^{}_{12}}$, while $c^\sigma_\tau$ can be obtained from
$\overline m^{}_1 c^\rho_\tau/\overline m^{}_2$ by making the replacement $t^{}_{12} \to -1/t^{}_{12}$.
These results can help us understand the numerical results:
(1) For $m^{}_1 \ll m^{}_2 \ll m^{}_3$, Eq. (\ref{2.3.6}) is simplified to
\begin{eqnarray}
c^\theta_\tau \simeq \frac{1}{2}  \;, \hspace{1cm}
c^\delta_\tau \simeq  \frac{\overline m^{}_2 c^{}_{12}s^{}_{12}} {m^{}_3 \bar s^{}_{13}}  \;,
\hspace{1cm} c^\rho_\tau \simeq  \frac{c^{}_{12} \bar s^{}_{13} }{s^{}_{12}}  \;.
\label{2.3.7}
\end{eqnarray}
(2) In the case of $m^{}_1 \simeq m^{}_2 \gg m^{}_3$, one will have
\begin{eqnarray}
c^\delta_\tau \simeq  \frac{1}{\rm T} \bar s^{}_{13} \;, \hspace{1cm}
c^\rho_\tau  \simeq  - \frac{[ m^{}_3+ \overline m^{}_1 (1- t^2_{12}) s^2_{13} ] \bar s^{}_{13}}{2 m^{}_3 t^{}_{12}} \;,
\label{2.3.8}
\end{eqnarray}
for $\rho^{(0)} = \sigma^{(0)}$, or
\begin{eqnarray}
c^\delta_\tau  \simeq  - \frac{8}{|r|} c^{}_{12}s^{}_{12}\bar s^{}_{13} \;, \hspace{1cm}
c^\rho_\tau  \simeq  \frac{8 [m^{}_3 + \overline m^{}_1 (c^2_{12}-s^2_{12})(1+ t^2_{12}) s^2_{13}]
c^3_{12} s^{}_{12} \bar s^{}_{13} } { m^{}_3 |r| }  \;,
\label{2.3.9}
\end{eqnarray}
for $\rho^{(0)} \neq \sigma^{(0)}$ together with $c^\theta_\tau \simeq -1/2$.
(3) When the case of $m^{}_1 \simeq m^{}_2 \simeq m^{}_3 \simeq m^{}_0$ is considered,
the coefficients approximate to
\begin{eqnarray}
& & c^\theta_\tau \simeq  \frac{2 m^2_0}{\Delta m^2_{31}} \;, \hspace{1.5cm}
 c^\delta_\tau \simeq \frac{2 m^2_0 [ {\rm T} r c^{}_{12} s^{}_{12} - 2 s^{2}_{13} ]}
{{\rm T} \Delta m^2_{31} \overline s^{}_{13}}  \;, \hspace{2.1cm}
c^\rho_\tau \simeq \frac{{\rm T} m^2_0 r c^{2}_{12} }{\Delta m^2_{31} \bar s^{}_{13}} \;; \nonumber \\
& & c^\theta_\tau \simeq \frac{2 m^2_0 c^2_{12}}{\Delta m^2_{31}}  \;, \hspace{1.32cm}
 c^\delta_\tau \simeq \frac{2 m^2_0 (r - 4  s^2_{13}) c^{}_{12} s^{}_{12} }
 { \Delta m^2_{21} \bar s^{}_{13}} \;, \hspace{2.3cm}
 c^\rho_\tau \simeq \frac{ 8 m^2_0 c^{3}_{12} s^{}_{12} \bar s^{}_{13}}{\Delta m^2_{21}} \;; \nonumber \\
& & c^\theta_\tau \simeq \frac{2 m^2_0 s^2_{12}}{\Delta m^2_{31}} \;, \hspace{1.3cm}
 c^\delta_\tau \simeq - \frac{2 m^2_0 ( r + 4 s^2_{13}) c^{}_{12} s^{}_{12} }
 { \Delta m^2_{21} \bar s^{}_{13}} \;, \hspace{2cm}
 c^\rho_\tau \simeq \frac{ 8 m^2_0 c^{3}_{12} s^{}_{12} \bar s^{}_{13}}{\Delta m^2_{21}} \;; \nonumber \\
& & c^\theta_\tau \simeq \frac{\Delta m^2_{31}}{8 m^2_0}  \;, \hspace{1.5cm}
 c^\delta_\tau \simeq - \frac{{\rm T} \Delta m^2_{21} c^{}_{12} s^{}_{12} + 2 \Delta m^2_{31} s^{2}_{13}}
{8 {\rm T} m^2_0 \overline s^{}_{13}}  \;, \hspace{1cm}
c^\rho_\tau \simeq \frac{\Delta m^2_{31} \bar s^{}_{13}}{8 m^2_0 t^{}_{12}} \;,
\label{2.3.10}
\end{eqnarray}
for $[\rho^{(0)}, \sigma^{(0)}] = [0, 0]$, $[\pi/2, 0]$, $[0, \pi/2]$ and $[\pi/2, \pi/2]$.
One can see that these approximation results agree well with the corresponding numerical results.

\begin{figure}[h]
\centering
\includegraphics[width=6.5in]{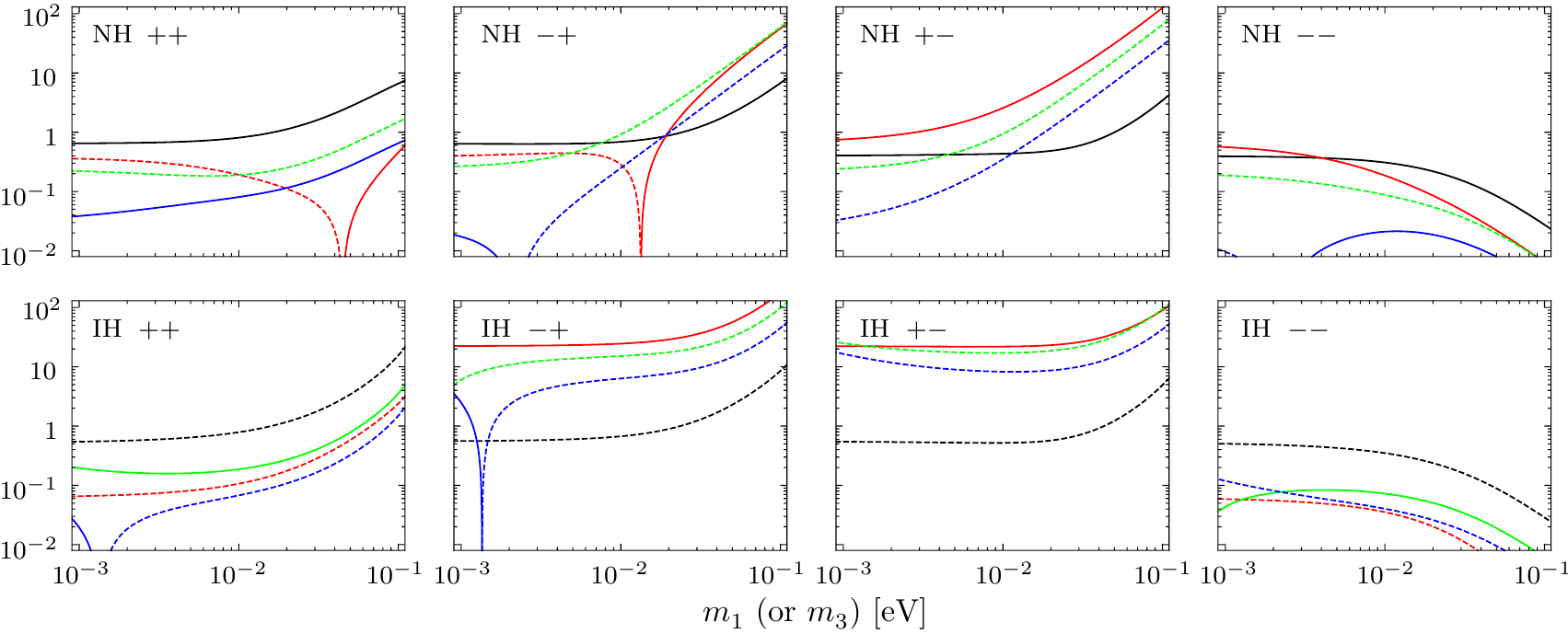}
\caption{The coefficients associated with $\Delta^{}_\tau$ ($c^{\theta}_{\tau}$ in black, $c^{\delta}_{\tau}$ in red,
$c^{\rho}_{\tau}$ in green and $c^{\sigma}_{\tau}$ in blue) against the lightest neutrino mass
$m^{}_1$ (or $m^{}_3$) in the NH (or IH) case for various combinations
of $\rho^{(0)}$ and $\sigma^{(0)}$ with $\delta^{(0)}=-\pi/2$. The signs $++, -+, +-$ and $++$
respectively stand for $[\rho^{(0)}, \sigma^{(0)}] = [0, 0], [\pi/2, 0], [0, \pi/2]$ and $[\pi/2, \pi/2]$.}
\end{figure}

\begin{table}[h]
\centering
\begin{tabular}{|p{3.5cm}<{\centering}|p{1cm}<{\centering}|p{3.5cm}<{\centering}|p{3.cm}<{\centering}|p{3cm}<{\centering}|} \hline
\backslashbox{[$\rho^{(0)}$, $\sigma^{(0)}$]}{$m^{}_1 [m^{}_3]$} & & 0.001 eV & 0.01 eV & 0.1 eV \\ \hline \hline
$[0, 0]$  & $c^{\theta}_{\tau}$ & 0.65 [$-$0.55] & 0.81 [$-$0.79] & 6.63 [$-$15.6]  \\ \hline
$[\pi/2, 0]$ & & 0.64 [$-$0.56] & 0.69 [$-$0.67] & 6.75 [$-$8.52] \\ \hline
$[0, \pi/2]$ & & 0.40 [$-$0.54] & 0.43 [$-$0.52] & 3.44 [$-$4.91] \\ \hline
$[\pi/2, \pi/2]$ & & 0.39 [$-$0.50] & 0.31 [$-$0.35]  & 0.03 [$-$0.03] \\ \hline \hline
$[0, 0]$  & $c^{\delta}_{\tau}$ & $-$0.36 [$-$0.07] & $-$0.19 [$-$0.11] & 0.46 [$-$2.39] \\ \hline
$[\pi/2, 0]$ & & $-$0.40 [22.4] & $-$0.28 [24.6] & 56.1 [180] \\ \hline
$[0, \pi/2]$ & & 0.76 [22.2] & 2.56 [21.9] & 125 [89.9] \\ \hline
$[\pi/2, \pi/2]$ & & 0.56 [$-$0.06] & 0.19 [$-$0.04]  & 0.01 [-0.00] \\ \hline \hline
$[0, 0]$  & $c^{\rho}_{\tau}$ & $-$0.22 [0.20] & $-$0.19 [0.19] & $-$1.49 [3.56]  \\ \hline
$[\pi/2, 0]$ & & $-$0.27 [$-$6.02] & $-$0.92 [$-$15.2] & $-$58.9 [$-$98.3]  \\ \hline
$[0, \pi/2]$ & & $-$0.24 [$-$25.0] & $-$0.94 [$-$17.1] & $-$65.3 [$-$93.9] \\ \hline
$[\pi/2, \pi/2]$ & & $-$0.19 [0.04] & $-$0.09 [0.07]  & $-$0.01 [0.01] \\ \hline \hline
$[0, 0]$  & $c^{\sigma}_{\tau}$ & 0.04 [0.02] & 0.08 [$-$0.07] & 0.65 [$-$1.50]  \\ \hline
$[\pi/2, 0]$ &  & 0.02 [2.47] & $-$0.25 [$-$6.34] & $-$24.1 [$-$45.2] \\ \hline
$[0, \pi/2]$ & & $-$0.03 [$-$16.5] & $-$0.35 [$-$8.26] & $-$29.2 [$-$42.9] \\ \hline
$[\pi/2, \pi/2]$ & & $-$0.01 [$-$0.12] & 0.02 [$-$0.04]  & 0.00 [$-$0.00]  \\ \hline
\end{tabular}
\caption{Some representative values of the coefficients associated with $\Delta^{}_\tau$
($c^{\theta}_{\tau}$, $c^{\delta}_{\tau}$, $c^{\rho}_{\tau}$ and $c^{\sigma}_{\tau}$ successively)
at $m^{}_1 (m^{}_3) = 0.001, 0.01$ and 0.1 eV in the NH (IH) case
for various combinations of $\rho^{(0)}$ and $\sigma^{(0)}$ with $\delta^{(0)}=-\pi/2$.}
\end{table}

\section{Summary}

To summarize, the $\mu$-$\tau$ reflection symmetry deserves particular attention as
it leads to the interesting results $\theta^{}_{23} = \pi/4$ and $\delta = \pm \pi/2$
(which are close to the current experimental data) as well as trivial Majorana
phases. Nevertheless, it is reasonable for us to consider the breaking of such a symmetry either
from the theoretical considerations (e.g., the RG running effect may provide a source
for the symmetry breaking) or on the basis of experimental results (e.g., the
newly-reported NOvA result disfavors the maximal mixing scenario at a 2.6$\sigma$
level). Consequently, we have performed a systematic study for the possible symmetry-breaking
patterns and their implications for the mixing parameters.

We first define some parameters measuring the symmetry-breaking strengths and then derive
an equation set relating them with the deviations of mixing parameters from the special
values taken in the symmetry context. By solving these equations in both a
numerical and analytical way, the sensitive strengths of mixing-parameter deviations to
the symmetry-breaking parameters for various neutrino mass schemes and
the Majorana-phase combinations are investigated in some detail. It turns out that $\Delta \theta$
is most sensitive to ${\rm R}^{}_2$ while $\Delta \delta$, $\Delta \rho$ and $\Delta \sigma$
to all the symmetry-breaking parameters. The coefficients for for $\Delta \delta$, $\Delta \rho$
and $\Delta \sigma$ are generally much greater than those $\Delta \theta$ in magnitude.
Furthermore, the coefficients tend to be magnified when the absolute neutrino mass scale increases
(in particular for the case of $m^{}_1 \simeq m^{}_2 \simeq m^{}_3$) and $\rho^{(0)}\neq \sigma^{(0)}$.
With these general results as guide,
one may easily find an appropriate specific way to break the $\mu$-$\tau$ reflection symmetry
so as to generate the required mixing-parameter deviations when necessary. Finally, as a unique
illustration, the general treatment is applied to the specific symmetry breaking induced by
the RG running effect.

\vspace{0.5cm}

\underline{Acknowledgments} \hspace{0.2cm} I would like to thank Professor Zhi-zhong Xing
for fruitful collaboration on the $\mu$-$\tau$ flavor symmetry. This work is supported
in part by the National Natural Science Foundation of China under grant No. 11605081.

\end{document}